\documentclass{article}

 \usepackage[main, final]{neurips_2026}

\usepackage[utf8]{inputenc} 
\usepackage[T1]{fontenc}    
\usepackage{hyperref}       
\usepackage{url}            
\usepackage{booktabs}       
\usepackage{amsfonts}       
\usepackage{nicefrac}       
\usepackage{microtype}      
\usepackage{xcolor}         

\usepackage{pifont}
\newcommand{\cmark}{\ding{51}}
\newcommand{\xmark}{\ding{55}}
\usepackage{amsmath}
\usepackage{amssymb}
\usepackage{amsthm}
\usepackage{graphicx}
\usepackage[table]{xcolor}
\usepackage{mathtools}
\usepackage{multicol, multirow}
\usepackage{float}
\usepackage{tcolorbox}
\usepackage{enumitem}
\usepackage{wrapfig}
\usepackage[table]{xcolor}
\definecolor{lightblue}{RGB}{235,245,255}
\definecolor{lightgreen}{RGB}{240,250,240}
\definecolor{lightgray}{RGB}{248,248,248}
\definecolor{lightorange}{RGB}{255,235,243}

\newtcolorbox{questionbox}{
    colback=blue!3,
    colframe=blue!50!black,
    boxrule=0.6pt,
    arc=2pt,
    left=6pt,
    right=6pt,
    top=6pt,
    bottom=6pt
}
\title{	
MindVLM: Neural-Grounded Visual Captioning via Subject-Aware Semantic Evidence Selection}

\author{
  Zixiang Yin$^\dag$, Yu-Ping Wang$^\ddag$, Zhengming Ding$^\dag$\thanks{Corresponding author. This work was supported by the U.S. National Science Foundation under Grant Nos. 2543694 and 2531809.}  \\
  $^\dag$Department of Computer Science, Tulane University \\
  $^\ddag$Department of Biomedical Engineering, Tulane University \\
  \texttt{\{zyin, wyp, zding1\}@tulane.edu}  \\
  \large\textcolor{magenta}{\nolinkurl{https://zachyin.com/mindvlm}}
}

\begin{document}

\maketitle

\begin{abstract}
Translating visual brain activity into natural language offers a powerful window into human cognition and neural semantic representation. Most prior brain-captioning approaches project fMRI signals into language space using paired supervision. However, the widely used \emph{Natural Scenes Dataset} (NSD) was built based only on image-fMRI pairs, while existing methods rely on \emph{MS-COCO} captions as indirect textual supervision. Critically, during fMRI acquisition, participants observe only cropped image regions, whereas the associated \emph{MS-COCO} captions describe full scenes. This mismatch creates a fundamental gap between neural representations and textual supervision, introducing an inherent \textbf{semantic loss} in NSD-based brain captioning. To mitigate this limitation, we reformulate brain captioning as \textbf{Neural-Grounded Visual Captioning}, where language is generated directly from visually grounded neural evidence rather than text supervision. We propose \textbf{MindVLM}, a framework that maps fMRI signals into subject-aware visual tokens and leverages a frozen vision-language model (VLM) to produce interpretable and personalized descriptions. Instead of point-wise regression, \textbf{MindVLM} focuses on extracting semantically relevant visual cues from neural activity, naturally accommodating partial observability and inter-subject variability. Extensive experiments on NSD show that \textbf{MindVLM} maintains, and often improves, caption quality across multiple VLM backbones, including LLaVA-v1.5, Qwen2.5-VL, and Gemma3, beyond intrinsic model priors. Despite requiring neither text supervision nor an LLM during training and operating under an image-grounded inference paradigm distinct from conventional captioning methods, \textbf{MindVLM} remains competitive with supervised approaches, suggesting that subject-specific neural grounding provides an efficient, scalable, and interpretable path toward faithful brain-to-language generation.
\end{abstract} 
\section{Introduction}
\label{sec:intro}

Understanding how the human brain encodes visual information is a central goal in cognitive neuroscience, offering fundamental insights into how humans perceive, interpret, and interact with the world~\cite{kay2023tasks}. Translating brain activity into natural language--commonly referred to as brain captioning--provides a concrete and expressive interface for probing these neural representations and their relationship to perception and language \cite{mathis2024decoding}. Recent progress in this line of work \cite{mindart,mindsa,mindgpt,yomind,neurocreat,mindllm,takagi2023improving} has demonstrated the feasibility of generating fluent descriptions from neural signals, largely driven by advances in large language models (LLMs) \cite{vicuna,gpt3,gpt2,gemma,qwen,llama}. Despite this progress, existing approaches implicitly assume that neural activity provides complete visual evidence for the target description--an assumption that does not hold in realistic experimental settings.


\begin{figure}[t]
  \centering
  \includegraphics[width=\columnwidth]{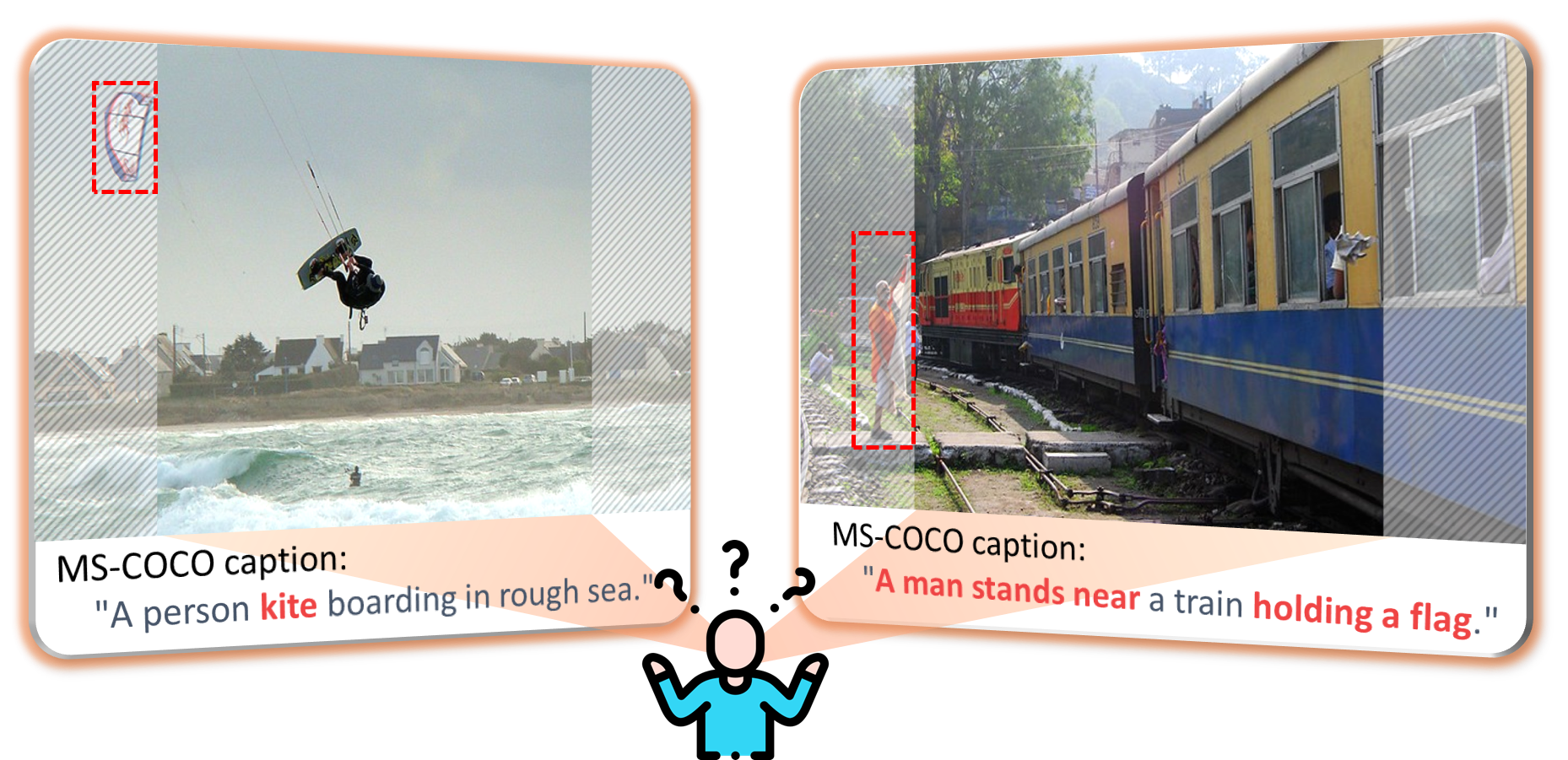}
  \vspace{-6mm}\caption{Examples illustrating semantic mismatches between cropped NSD images and their corresponding MS-COCO captions (see Figure~\ref{fig:semantic_loss}). Dimmed regions indicate portions of the original image removed during the NSD \cite{nsd} neural recording experiment. The \fcolorbox{red}{white}{red dashed boxes} mark objects that are physically \textbf{absent} from the cropped image but still \textbf{explicitly referenced} in the ground-truth captions, resulting in \emph{semantic loss}, where fMRI signals can only partially capture the full scene semantics. Dataset-level statistics are in Figure \ref{fig:semantic_loss}.}\label{fig:crop} \vspace{-5mm}
\end{figure}


In the widely-used Natural Scenes Dataset (NSD) \cite{nsd}, participants view \textbf{cropped regions} of MS-COCO images~\cite{mscoco} during fMRI acquisition, while textual supervision is derived from captions describing the \textbf{full scenes}. As illustrated in Figure~\ref{fig:crop}, this design introduces a fundamental mismatch between the visual evidence encoded in brain activity and the caption targets used for training, leading to what is referred to as \emph{semantic loss} in NSD. Directly using these captions as ground-truth supervision for brain captioning is not only computationally inefficient, but also risks pushing models to hallucinate visual content that was never actually observed in neural signals. Consequently, methods that attempt to reconstruct or regress toward complete visual representations---for example via point-wise alignment to image features or projection into LLM input space---may produce semantically inflated captions that exceed the information truly supported by the brain.

Moreover, fMRI signals primarily encode subject-specific perceptual and semantic representations \cite{baldassarre2012individual,phan2002functional}, reflecting individual differences in attention \cite{mangun1998erp}, feature sensitivity, and interpretation of visual scenes \cite{dubois2016building}. However, collecting paired fMRI-text data remains fundamentally challenging, as it is difficult to reliably record real-time textual interpretations precisely aligned with fMRI-correlated neural responses. As a result, existing brain captioning models largely overlook such individuality, producing generic descriptions that resemble image captions rather than language grounded in subject-specific neural evidence. At the same time, these approaches underutilize powerful vision-language models (VLMs), whose representations are directly grounded in visual semantics and well-suited for reasoning over perceptual neural signals \cite{doerig2025high}. These limitations highlight a critical gap in current methods and raise a central question:\vspace{-2mm}
\begin{questionbox}
\centering
\vspace{-2mm}\textit{How can we generate faithful, subject-aware captions by grounding descriptions only in the visual content that are truly supported by neural activity?}\vspace{-2mm}
\end{questionbox}
\vspace{-1mm}

To address this challenge, we introduce a new brain decoding paradigm: \textbf{Neural-Grounded Visual Captioning} via \emph{semantic evidence selection}. Instead of formulating brain captioning as direct fMRI-to-language translation, we redefine the task as identifying \emph{which visual semantics in a stimulus are supported by neural activity}. Specifically, we leverage the image as an intermediate semantic domain, treating spatial visual tokens as a structured semantic vocabulary over which neural evidence is inferred. Brain responses are used to select subject-aligned visual semantics, which are then composed into natural-language descriptions by a pretrained VLM. This image-grounded inference paradigm fundamentally departs from conventional fMRI-to-text generation by reframing brain decoding from \emph{text regression} to \emph{semantic evidence grounding}. As a result, it naturally handles partial visual observability, removes the need for costly subject-specific text annotations, enables efficient text-free training without LLM, and produces interpretable, subject-specific visual evidence maps that reveal individual perceptual focus.

Building on this framework, we introduce \textbf{MindVLM}, a subject-aware neural visual captioning model that captures the correspondence between brain responses and visual features. By learning to identify which visual tokens are supported by each subject’s unique neural signals, MindVLM constructs personalized visual evidence representations. These representations are then passed to a frozen vision-language model at inference time to generate descriptions that faithfully reflect the subject’s perceptual experience. Our contributions are summarized as follows:\vspace{-1mm}
\begin{itemize}[topsep=2pt, itemsep=2pt, parsep=0pt, leftmargin=*]
\item To the best of our knowledge, this is the first work to explicitly mitigate the \textbf{semantic loss} in NSD-based brain decoding, challenging the practicality of direct fMRI-to-language captioning due to the prohibitive cost of collecting reliable real-time subject-specific textual annotations.
\item We introduce \textbf{MindVLM}, an efficient, \textbf{text-free}, and subject-aware framework that discovers neural-supported visual tokens and leverages powerful pretrained VLMs to generate personalized, interpretable captions with strong semantic fidelity.
\item MindVLM can generalize across VLM architectures with distinct visual encoders, maintaining, and often improving, caption quality over each VLM baseline while remaining competitive with fully text-supervised methods on SPICE and CIDEr, despite relying on image-grounded inference rather than direct text regression, and achieving an order-of-magnitude lower training cost.
\begin{figure*}[t]
    \centering
    \includegraphics[width=\linewidth]{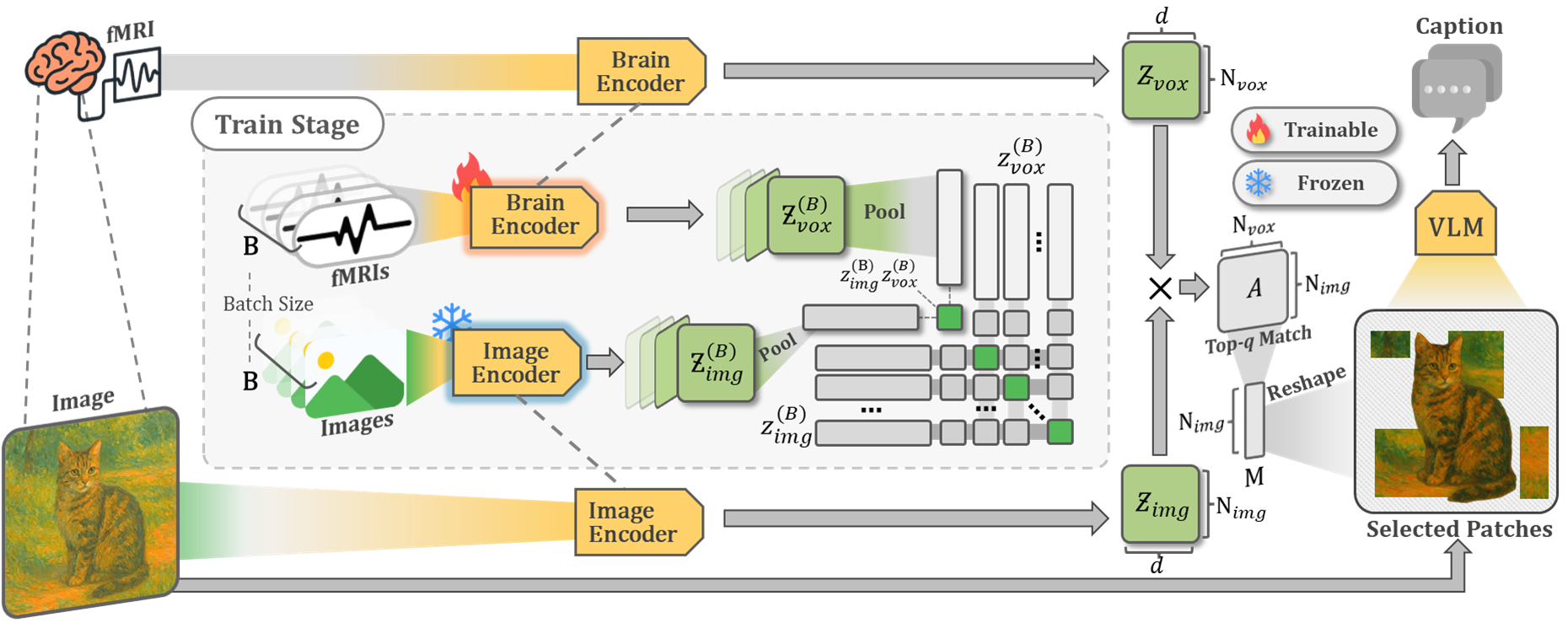}\vspace{-6mm}
    \caption{An overview of the proposed MindVLM neural-to-text framework. During training, a brain encoder is trained to align fMRI responses with the visual token representations extracted by a frozen VLM's image encoder. In inference, the most highly activated visual tokens are selected based on brain activity, and the frozen VLM attends to these selected tokens as semantic evidence to generate descriptive language output.}\vspace{-4mm}
    \label{fig:pipline}
\end{figure*}

\end{itemize}
\section{Related Work}
\label{sec:related_work}

\textbf{Neural-visual alignment.} Mapping fMRI into the representation space of pretrained vision or vision-language models is well established, with subject-specific or shared-subject encoders trained for image reconstruction and retrieval~\cite{mindtuner, brainclip,mindeye2,mindbridge}. These methods mostly aim to predict complete visual representations. UMBRAE~\cite{umbrae}, a unified brain decoding framework, maps fMRI into the visual-token space of a multimodal LLM without text supervision, synthesizing surrogate visual tokens that stand in for the image when captioning. In contrast, MindVLM aligns fMRI with visual tokens contrastively and, rather than synthesizing tokens or recovering the full image, selects the real visual tokens of the stimulus that each subject's neural response supports.

\textbf{Brain captioning.} Most brain captioning methods are trained with MS-COCO caption supervision~\cite{mindart,braincap,nv2lan}. MindGPT~\cite{mindgpt} distills CLIP semantics into a brain encoder and trains a GPT decoder on captions. MindLLM~\cite{mindllm} feeds fMRI as prefix tokens to a frozen LLM and trains its encoder end-to-end on text. Mind-SA~\cite{mindsa} masks image regions to build brain-compatible supervision for a frozen GPT-2 decoder. These captions describe the full MS-COCO scene, of which NSD subjects saw only a crop. MindVLM uses no text and no LLM during training, and its mask is the output, a per-subject evidence map, rather than a supervision signal. Our contribution is thus a shift from text-supervised decoding to Neural-Grounded Visual Captioning.
\section{Proposed Method}
\label{sec:method}

\subsection{Problem Formulation}

In a neural decoding dataset, each sample consists of a paired visual stimulus and its corresponding brain response. The visual stimulus is denoted by an RGB image $\mathbf{I} \in \mathbb{R}^{H \times W \times 3}$ with spatial resolution $H \times W$. The corresponding neural response $\mathbf{v} \in \mathbb{R}^L$ is flattened voxel-level fMRI activations recorded while a subject views the image. Here, $L$ refers to the number of voxels measured across visual and semantic cortices. Due to neuro-anatomical variability, $L$ naturally differs from subject to subject. In this work, we adopt a per-subject modeling paradigm, under which $L$ can be treated as fixed for a given subject.

\begin{wrapfigure}[19]{r}{0.56\columnwidth}
  \centering
  \vspace{-5mm}
  \includegraphics[width=0.56\columnwidth]{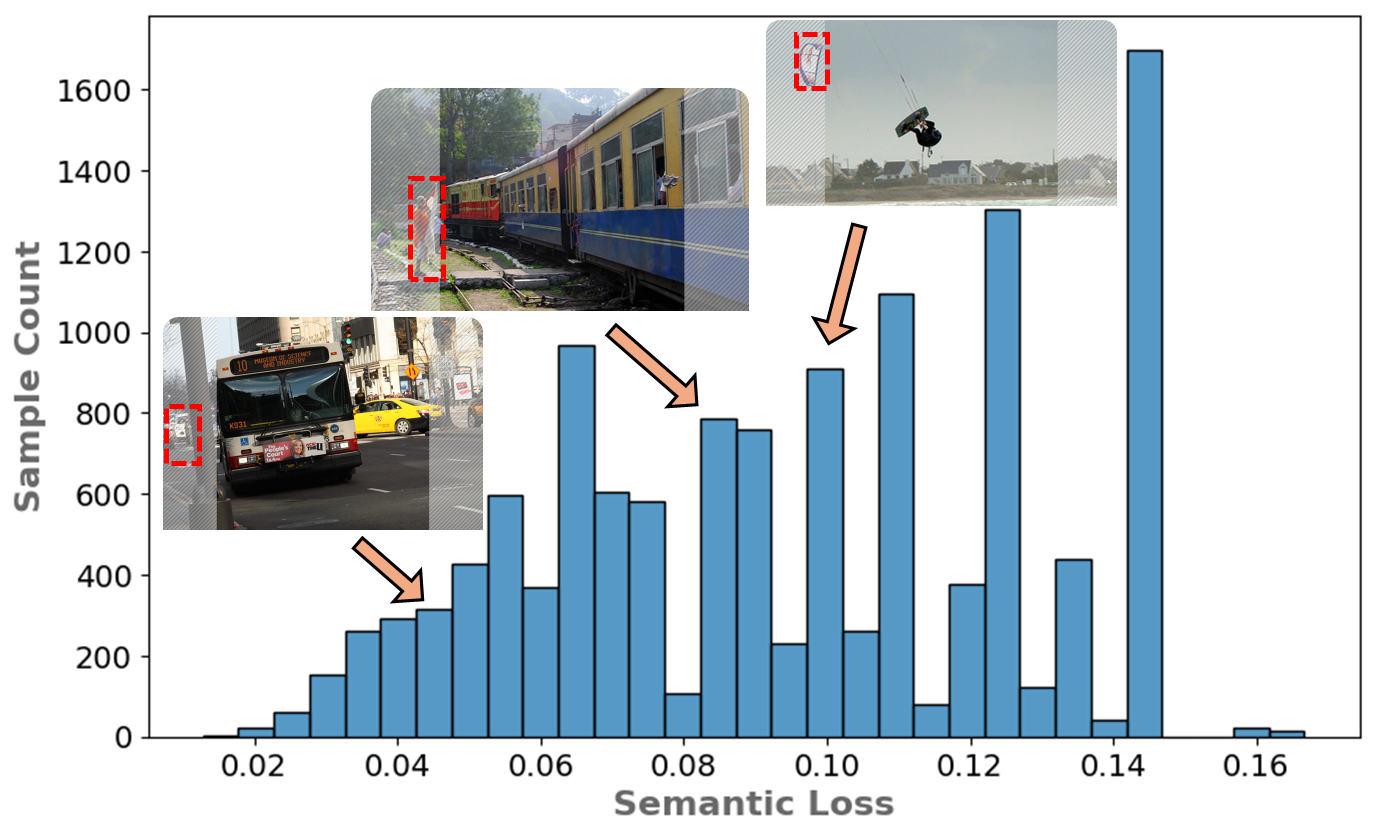}
  \vspace{-8mm}
  \caption{Distribution of sample counts with semantic loss in NSD. The histogram illustrates the semantic loss incurred when cropping original MS-COCO images to the NSD format. Semantic loss is defined as the fraction of objects in an image that lose more than $50\%$ of their total pixels during cropping \cite{nsd}. Notably, $12{,}901$ of $73{,}000$ images ($17.7\%$) exhibit semantic loss.}
  \label{fig:semantic_loss}
  \vspace{-6mm}
\end{wrapfigure}

Notably, many large-scale neural vision datasets derive their visual stimuli directly from established computer vision benchmarks. For example, the widely used NSD dataset \cite{nsd} samples images from MS-COCO \cite{mscoco}. Since MS-COCO provides multiple human-written, crowd-sourced captions per image, these paired image-text annotations naturally enable neural vision datasets to support the brain captioning task. Empirically, however, approximately 18\% of samples exhibit substantial semantic loss as shown in Figure~\ref{fig:semantic_loss}. A critical limitation is that NSD stimuli are often cropped versions of the original MS-COCO images, meaning that the associated captions may no longer accurately reflect the visual content presented during fMRI recording.

The goal of \textbf{Neural-Grounded Visual Captioning} is to design a module $\mathcal{M}$ that generates a textual description of a perceived image conditioned on its corresponding fMRI response. Formally, this task is defined as learning a mapping:
\begin{align}
    \mathcal{M}: \mathbb{R}^{H \times W \times 3} \times \mathbb{R}^{L} \rightarrow \mathcal{T},
\end{align}
where $\mathcal{T}$ denotes the space of natural language descriptions.


\subsection{Framework Overview}

Our framework is driven by three key insights: (1) \textbf{Semantic Loss}: fMRI signals capture only the visual content actually perceived by the subject, rather than the full semantic richness expressed in crowd-sourced captions; (2) \textbf{Personalization}: perception is subjective, as different individuals attend to different aspects of the same image, yet existing methods rely on subject-independent captions, making personalized annotations difficult to collect and enforcing uniform descriptions across subjects; and (3) \textbf{Visual Grounding}: pretrained VLMs provide structured, spatially grounded visual tokens whose representations naturally align with cortical visual features, offering a cognitively grounded and biologically plausible bridge for multimodal alignment.

For these reasons, we depart from conventional fMRI-to-text training pipelines. Instead, we construct a shared latent space where neural tokens derived from fMRI are aligned with visual tokens extracted from the associated image. Activated visual tokens then serve as the input to a frozen VLM for final caption generation (Figure \ref{fig:pipline}). During training, we first extract \textbf{spatial} neural features and \textbf{spatial} visual features, then aggregate them into global representations to perform batch-level contrastive matching, enabling robust learning of subject-specific neural-visual correspondences. During inference, we analyze spatial neural-visual similarity patterns to determine which visual tokens are most strongly reflected in the subject’s brain activity. A neuroscience-informed token selection mechanism then identifies the perceptually relevant visual tokens, which are passed to the frozen VLM to produce subject-specific captions that reflect individual’s interpretation of the image.


\subsection{Neural-Visual Mapping}

Given an image-fMRI pair $\mathbf{I}\in\mathbb{R}^{H\times W\times 3}$ and $\mathbf{v}\in\mathbb{R}^{L}$, our goal is to learn a shared latent space that captures the semantic correspondence between visual stimuli and brain activity patterns. We achieve this by first extracting latent spatial representations from both modalities.

\vspace{1mm}\noindent\textbf{Visual Encoding.} For the visual modality, we leverage the frozen image encoder $\mathcal{E}_\text{img}:\mathbb{R}^{H\times W\times3}\rightarrow\mathbb{R}^{N_\text{img}\times d_\text{img}}$ from the VLM of interest to obtain visual features $\mathbf{X}\in\mathbb{R}^{N_\text{img}\times d_\text{img}}$, providing rich semantic and spatial representations of the image:
\begin{align}
    \mathbf{X} := \mathcal{E}_\text{img}(\mathbf{I}) = (\mathbf{x}_1,\mathbf{x}_2,\cdots,\mathbf{x}_{N_\text{img}}) \in \mathbb{R}^{N_\text{img} \times d_\text{img}}.
\end{align}

\noindent\textbf{Neural Encoding.} For the neural modality, we adopt a trainable multi-head ViT-based \cite{vit} encoder $\mathcal{E}_\text{vox}:\mathbb{R}^{L}\rightarrow\mathbb{R}^{N_\text{vox}\times d_\text{vox}}$ to transform voxel-level fMRI signals into neural tokens $\mathbf{F}\in \mathbb{R}^{N_\text{vox} \times d_\text{vox}}$:
\begin{align}\label{eq:brain}
    \mathbf{F} := \mathcal{E}_\text{vox}(\mathbf{v})= (\mathbf{f}_1, \mathbf{f}_2, \dots, \mathbf{f}_{N_\text{vox}}) \in \mathbb{R}^{N_\text{vox} \times d_\text{vox}},
\end{align}
where $N_\text{vox}$ is the number of functional voxel tokens and $d_\text{vox}$ their latent dimensionality. Each voxel feature can be viewed as encoding activity patterns of local neuronal populations, reflecting the distributed semantic representation of visual information across the cortex.

\noindent\textbf{Learning Neural-Visual Shareable Space.} Considering the fact that the fMRI signal reflects the brain activity of the subject receiving visual stimulus at the image-level, it is important to align the two modalities globally. We therefore derive global representations by applying average pooling operation $\mathsf{Pool}(\cdot)$ over spatial dimensions:
\begin{align}
    \mathbf{x} = \mathsf{Pool}(\mathbf{X}) \in \mathbb{R}^{d_\text{img}} \quad\text{and}\quad \mathbf{f} = \mathsf{Pool}(\mathbf{F}) \in \mathbb{R}^{d_\text{vox}}.
\end{align}
This pooling operation mirrors the population-level coding observed in the visual cortex \cite{grill2004human}, where distributed patterns of voxel activity collectively represent the semantics of the images being viewed.

To enforce a global semantic consistency between brain and image representations, we project neural and visual global features into a shared embedding space of dimension $d$ using two linear heads $\mathbf{W}_\text{img}\in\mathbb{R}^{d_\text{img}\times d}$ and $\mathbf{W}_\text{vox}\in\mathbb{R}^{d_\text{vox}\times d}$:
\begin{align}\label{eq:g_proj}
    \mathbf{z}_\text{img} = \mathbf{W}_\text{img}^\top \mathbf{x}\in\mathbb{R}^d \quad\text{and}\quad 
    \mathbf{z}_\text{vox} = \mathbf{W}_\text{vox}^\top \mathbf{f}\in\mathbb{R}^d.
\end{align}
This shared space is expected to facilitate direct cross-modal matching and serves as the foundation for downstream neural-visual correspondence analysis.

\vspace{1mm}\noindent\textbf{Contrastive Learning.} To align neural and visual embeddings semantically in a shared space, we adopt a CLIP-style symmetric contrastive learning objective to jointly optimize $\mathbf{W}_\text{img}$, $\mathbf{W}_\text{vox}$, and $\mathcal{E}_\text{vox}$ in Eq.\eqref{eq:brain}. The goal is to maximize similarity between matched image-fMRI pairs while minimizing that between mismatched pairs within the same minibatch. The general form of a contrastive loss for a batch of $B$ query-target pairs $\{(\mathbf{u}^{(i)},\mathbf{g}^{(i)})|_{i=1}^B\}$ from two arbitrary populations $\mathcal{U}$ and $\mathcal{G}$ is:
\begin{align}
\mathcal{L}_{\mathcal{U}\rightarrow\mathcal{G}} = -\frac{1}{B}\sum_{i=1}^{B}\log\frac{\exp(\mathsf{cos}(\mathbf{u}^{(i)},\mathbf{g}^{(i)})/\tau)}{\sum\limits_{j=1}^{B}\exp(\mathsf{cos}(\mathbf{u}^{(i)}, \mathbf{g}^{(j)})/\tau)},
\end{align}
where $\mathsf{cos}(\cdot,\cdot)$ denotes cosine similarity and $\tau$ is a trainable temperature parameter controlling the sharpness of the similarity distribution. To build a high-quality mapping, we apply this contrastive loss in two symmetric directions with $\mathbf{z}_\text{img}$ and $\mathbf{z}_\text{vox}$: $\mathcal{L}_{\text{vox}\rightarrow\text{img}}$ for fMRI-to-image and $\mathcal{L}_{\text{img}\rightarrow\text{vox}}$ for the opposite. The final optimization object $\mathcal{L}_\text{align}$ is formulated as the average of the two losses:
\begin{align}\label{eq:loss}
    \mathcal{L}_\text{align}=\frac{1}{2}(\mathcal{L}_{\text{vox}\rightarrow\text{img}}+\mathcal{L}_{\text{img}\rightarrow\text{vox}}).
\end{align}
This bidirectional objective encourages mutual consistency between fMRI and image representations so that neural embeddings are capable of retrieving their corresponding visual counterparts, and vice versa. Through this symmetric alignment, the model learns a shared representational manifold where distributed cortical activity patterns correspond directly to semantically grounded visual features. 


\subsection{Subject-Aware Captioning}

Once the neural-visual alignment model is trained, it defines a shared embedding space that links voxel-level latent neural responses with spatially structured visual representations. This shared space naturally arises from the learning objective in Eq.\eqref{eq:loss}, in which both neural and visual features are globally pooled prior to alignment. Because the global token is obtained via average pooling over local tokens, it resides in the same embedding space as the local representations. This local-to-global design enables us to infer which image patches each subject attended to during perception, yielding interpretable and subject-specific attention maps directly from neural activity.

Given a \textbf{single} test pair of visual tokens $\mathbf{X}\in\mathbb{R}^{N_\text{img}\times d_\text{img}}$ and fMRI tokens $\mathbf{F}\in\mathbb{R}^{N_\text{vox}\times d_\text{vox}}$, both are projected into the shared space using the learned projection heads $\mathbf{W}_\text{img}\in\mathbb{R}^{d_\text{img}\times d}$ and $\mathbf{W}_\text{vox}\in\mathbb{R}^{d_\text{vox}\times d}$ similar to but slightly different from Eq.\eqref{eq:g_proj}, producing aligned \textbf{local} tokens $\mathbf{Z}_\text{img}$ and $\mathbf{Z}_\text{vox}$:
\begin{align}\label{eq:l_proj}
    \mathbf{Z}_\text{img} = \mathbf{X}\mathbf{W}_\text{img}\in\mathbb{R}^{N_\text{img}\times d}  \quad\text{and}\quad 
    \mathbf{Z}_\text{vox} = \mathbf{F}\mathbf{W}_\text{vox}\in\mathbb{R}^{N_\text{vox}\times d}.
\end{align}
A pairwise similarity matrix $\mathbf{A}$ is then calculated between the two modalities at local token level:
\begin{align}
\mathbf{A}=\mathbf{Z}_\text{vox}\mathbf{Z}_\text{img}^\top\in\mathbb{R}^{N_\text{vox}\times N_\text{img}}.
\end{align}
The matrix $\mathbf{A}$ captures the fine-grained neural-visual correspondence, where each $\mathbf{A}_{i,j}$ measures how strongly the $i$-th neural token (a localized cortical response) is associated with the $j$-th visual token. 

From a neuroscientific perspective \cite{jehee2011attention,shahdloo2022task}, the similarity between neural and visual tokens approximates subject-specific \textbf{attentional tuning}--that is, visual cortical voxels respond more strongly to features or regions that attract perceptual attention or encode behaviorally relevant semantics. However, neural responses are inherently distributed and overlapping: \textbf{multiple voxels may encode similar visual elements, while a single voxel may participate in multiple feature representations}.

To capture this many-to-many correspondence, we introduce a neuroscience-inspired selection mechanism. Instead of enforcing a one-to-one neural-visual mapping, we retain the top $q\%$ of neural-visual pairs based on their similarity scores in $\mathbf{A}$, yielding the filtered mask matrix $\mathbf{A}_q$. Similar to multiple instance learning, we compute a visual token-wise sum over $\mathbf{A}_q$ to obtain the final visual mask $\mathbf{m}_q \in \mathbb{R}^{N_\text{img}}$, where non-zero indices indicate visual tokens selected by aggregating diverse neural tokens. These indices highlight image regions most strongly supported by the neural activity, forming a semantic evidence mask that reflects the neural-driven attentional footprint of perception.

The selected visual tokens and the common text prompt ``\emph{Please describe the image as simply as possible.}'' are then passed to a frozen VLM, which interprets subject-aware fMRI-activated visual tokens to generate individualized textual descriptions of the perceived content. Importantly, the image itself is never altered: the vision encoder always encodes the complete, unmodified stimulus, so every visual token carries its full-image representation, and selection only restricts, through an attention mask, which of these tokens the language model attends to. Thus, our model decodes fMRI signals not by directly regressing to text, but by inferring which visual evidence the subject attended to and using the VLM to translate that evidence into language.    
\section{Experiments}
\label{sec:results}

\subsection{Experimental Setup}

We evaluate our MindVLM on the Natural Scenes Dataset (NSD)~\cite{nsd}, the largest publicly available fMRI dataset. NSD captures brain activity from 8 human subjects as they viewed natural scenes from MS-COCO~\citep{mscoco}. Each subject passively viewed a set of 10,000 images, each presented for 3 seconds and repeated three times. Subjects were instructed to maintain central fixation throughout the experiment. Among these images, 1,000 are shared across all subjects, while the remaining 9,000 are unique to individual subjects. Due to data availability constraints, we only train and test on four subjects: subj01, subj02, subj05, and subj07. Following standard protocols established in prior NSD studies~\cite{data_prep1,brainclip,mindeye1,mindeye2,data_prep2,imind}, we used standardized train/test splits and averaged fMRI activations over repetitions for each image within each subject. This pre-processing resulted in 36,000 training samples (fMRI-image pairs) and 4,000 test samples. All models are then trained and evaluated on a per-subject basis. Check Appendix \ref{subsec:impl} for implementation details.

\subsection{Neural-Grounded Visual Captioning across VLM Architectures}

We introduce \textbf{Neural-Grounded Visual Captioning (NGVC)} as a new paradigm for brain decoding that explicitly mitigates the difficulty of collecting personalized fMRI-text pairs. Unlike traditional brain captioning methods that directly map neural activity into language space, NGVC leverages neural responses to identify semantically supported visual tokens from the test image, which are then passed to a frozen VLM for caption generation. Crucially, the brain encoder is trained to align with the VLM's \emph{visual encoder space} rather than its language space, eliminating the need for costly subject-specific text supervision while enabling personalized and semantically grounded captioning.

To evaluate generality across architectures, we train separate brain encoders aligned with three commonly-used VLMs featuring different visual backbones: LLaVA-v1.5-7B (CLIP), Qwen2.5-VL-7B-Instruct (customized ViT), and Gemma3-4B (SigLIP). For each model, we compare MindVLM against the base VLM operating on the full image, which serves as a strong baseline reflecting the model’s inherent visual understanding without neural guidance. We report results on 8 standard captioning metrics, including n-gram-based measures (BLEU 1-4 \cite{bleu}, ROUGE \cite{rouge}) and semantic metrics (METEOR \cite{meteor}, CIDEr \cite{cider}, SPICE \cite{spice}). Results for all four subjects are shown in Table~\ref{tbl:vlm_compat}, with $q=70$ for all three backbones. This high-retention setting keeps most visual tokens and filters out only those least supported by the neural response, so comparing MindVLM with the base VLM on the full image tests whether caption quality holds when part of the visual input is withheld.

\begin{table}[t]
  \caption{Neural-Grounded Visual Captioning evaluation across VLMs using different visual encoders for four NSD subjects at $q=70$. \textbf{Bold} marks the best value per column within a backbone.}\vspace{-2mm}
  \label{tbl:vlm_compat}
  \centering
  \setlength{\tabcolsep}{3pt}
  \resizebox{\textwidth}{!}{%
  \begin{tabular}{@{}ccccccccccc@{}}
    \toprule
    & & & \multicolumn{4}{c}{\textbf{BLEU-n Reference}} & \multicolumn{4}{c}{\textbf{Content \& Semantic Agreement}} \\
    \cmidrule(lr){4-7}\cmidrule(lr){8-11}
    VLM (Encoder) & Method & Subject & B@1 & B@2 & B@3 & B@4 & ROUGE & METEOR & CIDEr & SPICE \\
    \midrule
    \multirow{5}{*}{\shortstack[c]{Qwen2.5-VL-7B~\cite{qwen}\\\small(Customized ViT)}} & Original & -- & 49.52 & 32.16 & 20.93 & 13.63 & 39.84 & 21.92 & 46.13 & 14.95 \\
     & \multirow{4}{*}{MindVLM} & subj01 & 49.77 & 32.42 & 21.13 & 13.70 & 40.08 & 22.09 & 46.52 & 14.94 \\
     &  & subj02 & \textbf{50.17} & \textbf{32.85} & \textbf{21.49} & 13.91 & \textbf{40.32} & 22.18 & \textbf{47.28} & \textbf{15.35} \\
     &  & subj05 & 49.86 & 32.37 & 20.99 & 13.66 & 40.05 & 22.07 & 46.58 & 14.91 \\
     &  & subj07 & 49.91 & 32.59 & 21.36 & \textbf{14.00} & 40.21 & \textbf{22.20} & 47.17 & 15.07 \\
    \midrule
    \multirow{5}{*}{\shortstack[c]{LLaVA-v1.5-7B~\cite{llava}\\\small(CLIP ViT-L/14)}} & Original & -- & 55.80 & 40.14 & 27.31 & 18.11 & 47.40 & 27.42 & 60.61 & 21.59 \\
     & \multirow{4}{*}{MindVLM} & subj01 & 56.21 & 40.55 & 27.65 & 18.35 & 47.53 & 27.42 & 62.06 & 21.61 \\
     &  & subj02 & 55.86 & 40.47 & \textbf{27.72} & \textbf{18.50} & 47.65 & 27.38 & 61.64 & 21.52 \\
     &  & subj05 & 55.68 & 40.11 & 27.32 & 18.26 & 47.38 & \textbf{27.43} & 59.51 & 21.50 \\
     &  & subj07 & \textbf{56.51} & \textbf{40.58} & 27.51 & 18.24 & \textbf{47.68} & 27.39 & \textbf{62.98} & \textbf{21.70} \\
    \midrule
    \multirow{5}{*}{\shortstack[c]{Gemma3-4B~\cite{gemma}\\\small(SigLIP)}} & Original & -- & 68.35 & 49.16 & 33.33 & 21.80 & \textbf{49.26} & \textbf{25.37} & \textbf{92.12} & 20.95 \\
     & \multirow{4}{*}{MindVLM} & subj01 & 68.04 & 48.52 & 32.78 & 21.42 & 48.59 & 25.13 & 91.19 & 20.92 \\
     &  & subj02 & \textbf{68.39} & \textbf{49.44} & \textbf{33.58} & \textbf{21.88} & 49.20 & 25.18 & 92.11 & \textbf{21.08} \\
     &  & subj05 & 67.68 & 48.38 & 32.66 & 21.30 & 48.68 & 25.04 & 90.56 & 20.69 \\
     &  & subj07 & 68.18 & 49.00 & 33.27 & 21.69 & 49.16 & 25.26 & 91.78 & 20.78 \\
    \bottomrule
  \end{tabular}}\vspace{-5mm}
\end{table}

\noindent\textbf{Captioning Performance.} MindVLM is comparable to the base VLM across all three architectures and often outperforms it, even though the language model attends only to the selected visual tokens. For Qwen2.5-VL, MindVLM exceeds the base VLM in 30 of 32 subject--metric pairs, missing only SPICE for subj01 and subj05 by at most 0.04. For LLaVA, neural-guided selection yields gains on most metrics for three of the four subjects, with CIDEr increasing from 60.61 to as high as 62.98. For Gemma3, a substantially stronger base model, MindVLM trails the base VLM slightly for subj01, subj05, and subj07 and exceeds it for subj02 on BLEU@1-4 and SPICE, reflecting the diminishing room for improvement as base model quality increases. These results indicate that neural-guided token selection preserves or improves caption quality across CLIP, ViT, and SigLIP encoder families, suggesting that the paradigm generalizes across architectural choices.

\begin{wraptable}[6]{r}{0.5\textwidth}
\centering\vspace{-6mm}
\caption{Reference-free Validation.}
\label{tbl:external_score}

\begin{tabular}{ccccc}
    \toprule
    Scorer & COCO & Ours & $\Delta$ [95\% CI] \\
    \midrule
    CLIP   & 65.26 & \textbf{66.85} & $+1.59$ [1.23, 1.95] \\
    SigLIP & 28.83 & \textbf{29.61} & $+0.78$ [0.44, 1.13] \\
    \bottomrule
  \end{tabular}
\end{wraptable}

\noindent\textbf{Reference-Free Validation.} To further validate that captions generated from the selected tokens remain faithful to the presented image, we conduct a reference-free evaluation with two image-text models from independent families as external scorers: CLIP score~\cite{clipscore} and SigLIP~\cite{siglip}. Both score the semantic alignment between a caption and an image. We score each MS-COCO and MindVLM caption directly against the NSD stimulus the subject viewed. Each of the 1,000 test images contributes one paired difference, with MindVLM's score averaged over the four subjects.

Table~\ref{tbl:external_score} reports the mean difference with a 95\% confidence interval (CI) from a paired bootstrap (10,000 resamples). Under both scorers, MindVLM-LLaVA captions align with the presented stimulus more closely than MS-COCO captions, which were written for the uncropped images. Grounding generation in selected tokens thus yields captions more faithful to what the subject saw than human references written for the full scene, even though the LLM attends to only part of the visual input.










\vspace{-2mm}
\subsection{Semantic Decoding Quality: Reference to Brain Captioning Methods}
To situate MindVLM within the broader brain captioning literature, we report the performance of MindVLM-LLaVA at $q=35$ on the standard NSD evaluation protocol alongside established fMRI-to-text methods. Here, $q = 35$ is a conservative choice to maximize the fairness of reference. See detailed analysis in Section~\ref{subsec:q_anaysis} and \ref{subsec:q_analysis_app}. We emphasize that this comparison is provided as \textbf{a reference point, not a direct benchmark}. Traditional brain captioning methods perform pure fMRI-to-text decoding with no image access at inference, whereas MindVLM operates under the NGVC paradigm--using fMRI to select neural-supported visual tokens from the test image, which are then passed to a frozen VLM. These are fundamentally different inference settings, and the comparison is intended to characterize the semantic quality of neural grounding rather than to establish ranking. 

Despite requiring no text supervision and operating under a different inference paradigm, MindVLM achieves strong performance on semantic-oriented metrics in Table~\ref{tb:brain_cap}. On SPICE, we substantially exceed all methods, including fully text-supervised ones, suggesting that grounding captions in neural-supported visual evidence captures scene-level semantics, object relations, and compositional meaning more faithfully than direct text regression. On ROUGE, MindVLM also leads across all methods, indicating strong coverage of content-relevant words even without paired caption training. On n-gram metrics, MindVLM trails text-supervised methods, which is expected, as they reward lexical similarity to ground-truth captions, something inherently unattainable without text supervision. We also find no evidence that MS-COCO leakage into LLaVA inflates these metrics (Appendix~\ref{app:leakage}).

One additional finding is worth noting. MindVLM-clean (evaluated on the subset of test samples without semantic loss) consistently outperforms MindVLM on the full test set across all metrics, with the gap most pronounced on CIDEr. This directly quantifies the partial visual observability issue: when neural signals and caption targets are well-aligned, semantic grounding improves measurably, confirming that the problem identified in Section~\ref{sec:intro} has real and measurable consequences for decoding quality. Section~\ref{subsec:loss} analyzes this relationship in further detail. These results point to a broader insight about what text supervision actually provides in brain captioning: (1) text-supervised models excel at \textbf{reproducing} captions, but at the cost of learning dataset-level correlations that may exceed what the subject actually perceived. (2) MindVLM, by grounding decoding in neural-supported visual evidence rather than crowd-sourced text, trades lexical precision for semantic faithfulness--a trade-off that favors the underlying goal of brain captioning as a window into individual perception.
\begin{table}[t]
\centering
\caption{Reference comparison between MindVLM-LLaVA at $q=35$ and the conventional brain captioning methods on NSD, averaged across subj01, subj02, subj05, and subj07. ``\textbf{Text Sup.}'' column marks methods trained with MS-COCO caption supervision. MindVLM uses no text supervision.}\vspace{-2mm}
\resizebox{\textwidth}{!}{%
\begin{tabular}{lc>{\columncolor{gray!4}}c
                  >{\columncolor{yellow!5}}c
                  >{\columncolor{gray!6}}c
                  >{\columncolor{pink!5}}c
                  >{\columncolor{green!4}}c
                  >{\columncolor{red!3}}c
                  >{\columncolor{orange!5}}c
                  >{\columncolor{blue!4}}c }
\toprule
                 & & \multicolumn{4}{c}{\textbf{BLEU-n Reference}} & \multicolumn{4}{c}{\textbf{Content \& Semantic Agreement}} \\
\cmidrule(lr){3-6} \cmidrule(lr){7-10}
\textbf{Methods} & \textbf{Text Sup.} & \textbf{B@1} & \textbf{B@2} & \textbf{B@3} & \textbf{B@4} & \textbf{ROUGE} & \textbf{METEOR} & \textbf{CIDEr} & \textbf{SPICE} \\
\midrule
 BrainCap \cite{braincap}                 & \cmark & 54.82 & 35.14 & 21.96 & 14.04 & 40.39 & 16.24 & 38.86 & 8.84  \\
 MindLLM \cite{mindllm}                   & \cmark & 56.23 & 36.60 & 23.66 & 15.91 & 41.56 & 16.16 & 39.97 & 8.82  \\
 Neuro2Language \cite{nv2lan}             & \cmark & 57.19 & 37.17 & 23.78 & 15.85 & 36.67 & 18.60 & 49.51 & 12.39 \\
 UMBRAE \cite{umbrae}                     & \xmark & 57.38 & 37.98 & 25.18 & 17.11 & 41.99 & 18.27 & 51.68 & 11.82 \\
 MindGPT \cite{mindgpt}                   & \cmark & 56.91 & 36.95 & 23.41 & 15.87 & 38.41 & 20.04 & 43.56 & 10.97 \\
 MindArt \cite{mindart}                   & \cmark & 57.26 & 38.62 & 25.17 & 17.71 & 42.86 & 21.93 & 45.95 & 12.27 \\
 Mind-SA \cite{mindsa}                    & \cmark & 59.76 & 39.81 & 27.05 & 18.86 & 42.72 & 38.08 & 54.03 & 12.25 \\
\midrule
 MindVLM (Ours)$^*$                       & \xmark & 54.73 & 39.35 & 26.64 & 17.63 & 46.41 & 27.13 & 58.25 & 21.05 \\
 MindVLM (Ours)                           & \xmark & 54.03 & 38.80 & 26.21 & 17.32 & 46.00 & 26.90 & 54.82 & 21.04 \\
\bottomrule
\multicolumn{10}{l}{\small * indicates the performance of our model on the subset of NSD test set without semantic loss. 
Detailed in Section \ref{subsec:loss}.
} \vspace{-4mm}
\end{tabular}
}
\label{tb:brain_cap}
\end{table}

\subsection{Role and Choice of Token Selection $q$}
\label{subsec:q_anaysis}

\begin{wrapfigure}[10]{r}{0.6\columnwidth}
  \centering\vspace{-4mm}
  \includegraphics[width=0.6\columnwidth]{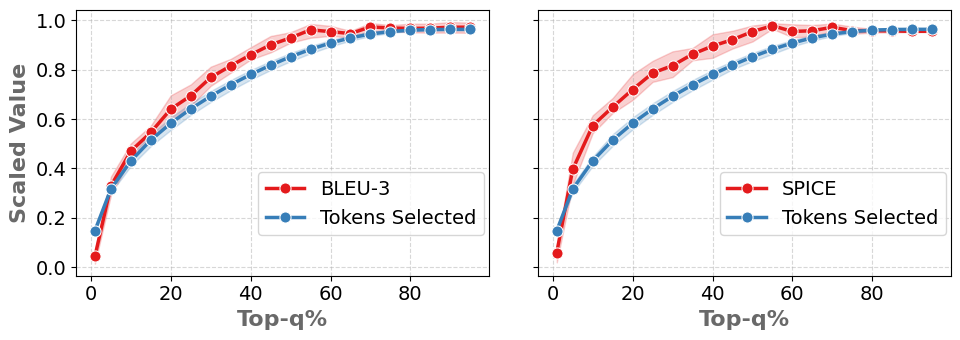}\vspace{-8mm}
  \caption{Scaled BLEU-3 and SPICE metrics with visual tokens ratio \emph{vs.} top $q\%$ selected fMRI-image token pairs.}\label{fig:metric_q}
\end{wrapfigure}

We present an analysis of how $q$ influences the captioning performance. We report two representative metrics (other metrics provided in Appendix \ref{subsec:cap_res_all} and \ref{subsec:metric_q}): BLEU-3, capturing lexical-level (n-gram) similarity, and SPICE, which quantifies semantic consistency between predicted and reference captions. As shown in Figure \ref{fig:metric_q}, both the scores and the proportion of selected visual tokens are min-max normalized to emphasize their relative evolution as $q$ increases. BLEU-3 and SPICE exhibit a general increase with $q$, confirming that selecting more fMRI-visual correspondences tends to improve decoding quality. Quantitatively, performance rises sharply up to about the top 40-55\% pairs, where all subjects on average already achieve over 90\% of their peak BLEU-3 and SPICE values. Beyond this point, the curves flatten or even slightly decline. This suggests that additional tokens contribute little or may even introduce noise. The rise with $q$ also indicates that the frozen VLM's language prior alone does not produce the captions.

From a neuroscience perspective, this behavior aligns with the notion of redundant or weakly informative voxels in visual cortex: while many voxels encode overlapping visual semantic content, only a subset carries strong, discriminative signals aligned with the textual semantics. Excessive tokens may reintroduce background information, diluting semantic alignment. 

\subsection{Captioning by Subject-Specific Perception}\label{subsec:ssperception}
\begin{figure}[b]
  \centering\vspace{-7mm}
  \includegraphics[width=\textwidth]{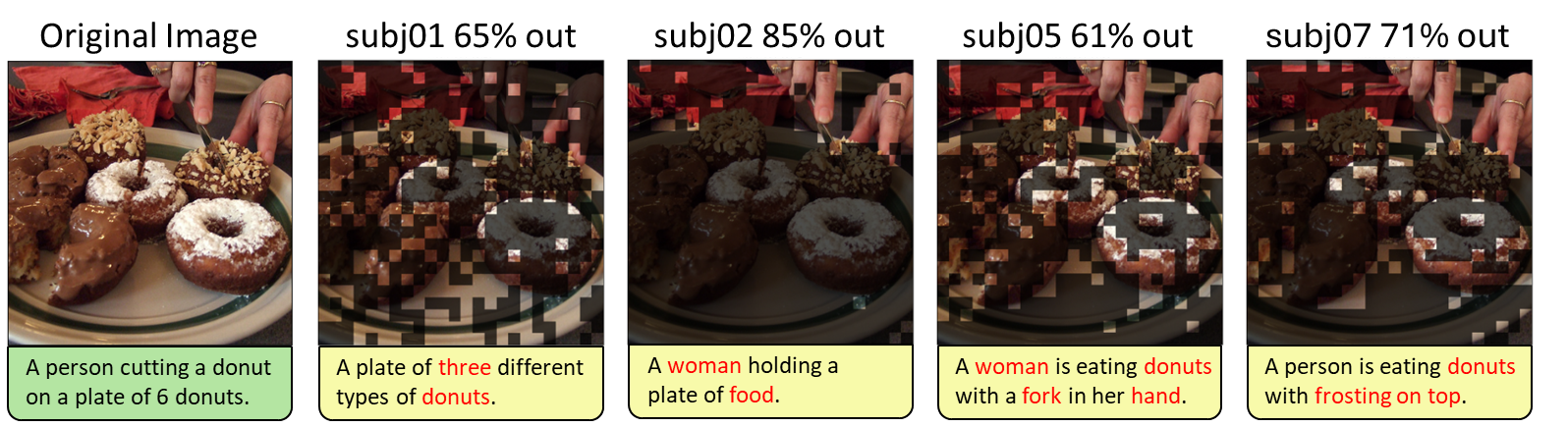}
  \vspace{-6mm}\caption{Language generation based on personalized perception at $q=5$ for all subjects.}\label{fig:vis_text} \vspace{-3mm}
\end{figure}
Figure \ref{fig:vis_text} illustrates how the same visual stimulus leads to distinct, subject-specific perceptual interpretations that ultimately shape the generated captions. Although we use a fixed quantile threshold $q = 5$, the proportion of preserved tokens (``XX\% out'') varies across individuals, indicating differences in neural selectivity and revealing what their brain considers most relevant or salient percepts. Importantly, captions generated from these subject-specific tokens do not resemble generic image captions. Instead, they capture unique visual information encoded by each individual, highlighting the well-documented inter-individual variability in high-level visual cortex \cite{baldassarre2012individual}.

Specifically, \textbf{subj02} preserves very few tokens (85\% masked), particularly around object boundaries. This suggests a coarse, global percept lacking fine-grained object details. Correspondingly, the generated caption--``\textit{a woman holding a plate of food}.''--remains semantically broad and avoids mentioning items such as ``donuts'' or ``fork,'' which are either absent from the subject's neural representation or not prioritized during perception. In contrast, \textbf{subj05} retains the largest proportion of tokens (only 61\% masked), with preserved regions concentrated around the action and utensil. This indicates a higher-fidelity perceptual representation. As a result, the generated caption becomes detail-rich and action-specific (``\textit{eating ... with a fork in her hand}''), showing that MindVLM reconstructs the fine-grained semantics embedded in its neural patterns. \textbf{Subj07} exhibits selective retention around the pastries, preserving tokens that emphasize high-contrast texture regions. This subject-specific attentional bias results in a caption that uniquely includes ``\textit{frosting on top}.'' Others whose neural encodings did not emphasize these texture cues do not mention this detail, further demonstrating that the decoded text reflects personalized perceptual biases rather than a caption-driven ground truth.

Together, these results (more in Appendix \ref{subsec:vis_more}) demonstrate that the decoded captions are personalized semantic reconstructions, grounded in each individual's neural representation of the visual stimulus. Differences in information retention, spatial focus, and feature selectivity translate directly into systematically different linguistic outputs. Thus, unlike existing approaches that regress toward an average or caption-conditioned description, our paradigm leverages subject-specific cortical encoding to reconstruct what each person actually perceived. This supports the claim that neural-to-text decoding should be framed as subject-dependent semantic reconstruction, rather than text prediction based solely on crowd-sourced caption annotations.

\subsection{Paradigm Efficiency}

One key advantage of MindVLM over conventional captioning models lies in its \textbf{data efficiency} and \textbf{lightweight training} design. Our approach eliminates the dependence on text-based supervision, which is both scarce and difficult to obtain. Most competing approaches rely on using images as an indirect bridge between brain signals and language, often leading to cross-modal misalignment.

\begin{wrapfigure}[12]{r}{0.7\columnwidth}
  \centering\vspace{-4mm}
  \includegraphics[width=0.7\columnwidth]{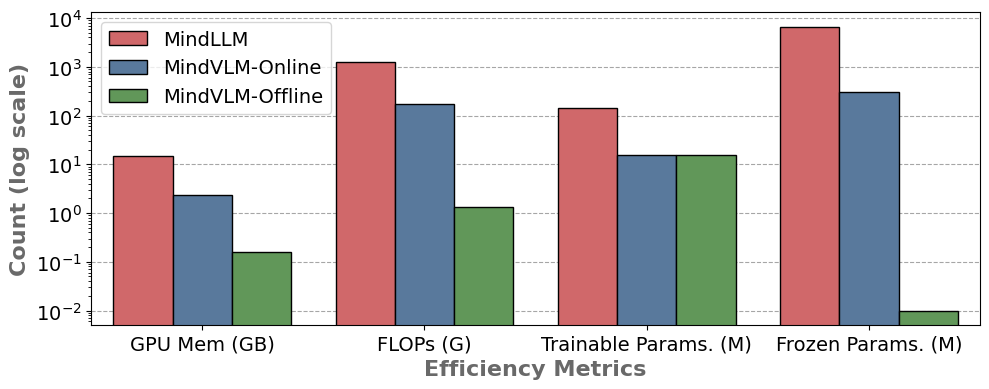}\vspace{-7mm}
  \caption{Model efficiency comparison against MindLLM.}\label{fig:effi}
\end{wrapfigure}

In addition to reduced data requirements, MindVLM is significantly more \textbf{computationally efficient}. Because our training objective involves only neural-visual contrastive alignment, no LLM is required during optimization, resulting in dramatically lower resource consumption. To quantitatively assess training efficiency, we benchmark MindVLM against the recent MindLLM model \cite{mindllm} across four common metrics: a) peak GPU memory (in GB and with a single sample), b) floating-point operations (FLOPs) in gigaflops (G), trainable \& frozen parameters in million (M). In Figure \ref{fig:effi}, we report two configurations: MindVLM-Online, where the CLIP visual encoder operates within the training loop, and MindVLM-Offline, where image features are precomputed and loaded directly during training.

The results highlight a substantial computational advantage. MindVLM-Online and Offline require only \textbf{10\%} and \textbf{1\%} of the GPU memory used by MindLLM, respectively. In terms of computational cost, MindLLM consumes about \textbf{10$\times$} more FLOPs than the online variant and over \textbf{1,000$\times$} more than the offline variant. Both MindVLM configurations maintain a lightweight architecture with only \textbf{15M} trainable parameters, compared to MindLLM's \textbf{141M}. Moreover, due to its text-supervised training paradigm, MindLLM must load an LLM during optimization, resulting in a substantial number of frozen parameters. In contrast, MindVLM can operate entirely in an offline setting, where nearly no frozen parameters are required. Importantly, these gains in efficiency do not come at the cost of performance. As shown in Table \ref{tb:brain_cap}, MindVLM surpasses MindLLM across nearly all metrics--except for B\@1--despite being trained without any text supervision.\vspace{-2mm}



\section{Conclusion}
\label{sec:conclusion}
We presented MindVLM, a subject-aware framework for Neural-Grounded Visual Captioning that reframes brain decoding as semantic evidence selection under partial visual observability. Rather than mapping fMRI signals directly to text, MindVLM aligns neural activity with spatial visual tokens and uses a frozen VLM to generate captions grounded only in the neural-supported visual evidence of each subject. MindVLM performs comparably to, and often better than, the base VLM across three architectures. In addition, it remains competitive with text-supervised brain captioning methods on semantic metrics despite requiring no text supervision--with remaining performance gaps concentrated on lexical-overlap metrics that favor models trained to reproduce captions. These results indicate the promise of neural-guided visual grounding as an efficient, interpretable, and cognitively faithful pathway for personalized decoding.

\textbf{Limitations.} MindVLM is a preliminary step toward modeling subject-specific perception. NSD records neither what each subject attended to nor how they would describe the image. The subject-specific selections are therefore not yet validated against independent measures of what each subject perceived, and our evidence for them is observational. Testing them directly requires a dataset in which subjects describe what they saw, ideally under free viewing. MindVLM trains a separate encoder for each subject and each visual backbone. Its captions are bounded by what the frozen VLM can express, and it is evaluated only on NSD. Future work could extend MindVLM to multi-subject training and use its neurally grounded captions to supervise an fMRI-to-text model.
\newpage

\bibliographystyle{plain}
\bibliography{main}


\newpage
\appendix
\section{Appendix}
\subsection{Related Works}
\label{subsec:rw}

Brain-to-language decoding spans a broad set of tasks and neural modalities. Early and ongoing lines of work study how to decode fMRI or other neural signals during story listening \cite{tang2023semantic, unicorn, lebel2023natural, antonello2023scaling, mapguide, lubrain} or reading visually presented text in a sentence-by-sentence manner \cite{pereira2018toward, fedorenko2016neural, schrimpf2021neural, ye2025generative, zou-etal-2022-cross}.
Our work focuses on a specific subproblem within this landscape: brain captioning, where the goal is to generate natural-language descriptions of the semantic content of visual stimuli viewed by subjects inside the scanner. This setting is enabled primarily by the Natural Scene Dataset (NSD) \cite{nsd}, with images from MS-COCO \cite{mscoco}, and thus provide rich human-written captions for evaluation.

\subsubsection{Modality Alignment}
A substantial body of prior work investigates how to align neural activity with visual or vision-language representations as a foundation for downstream decoding. These efforts differ in motivation, supervision, and target tasks, but generally aim to improve the fidelity or robustness of neural-visual correspondence.

Several works focus on learning subject-unified representations that generalize across individuals. UMBRAE \cite{umbrae} proposes a universal brain encoder that maps subject-specific fMRI signals into a shared latent space corresponding to visual tokens of a multimodal model. Its design emphasizes cross-subject consistency and supports weakly supervised adaptation to new subjects, enabling scalable neural decoding without text supervision during training. BrainOT \cite{brainot} addresses limitations of point-wise regression by reformulating neural-visual alignment as a distribution-to-distribution matching problem using optimal transport, improving robustness over traditional mean-squared-error objectives when aligning fMRI signals with visual features.

Another line of work prioritizes neural-to-image reconstruction as the primary objective. MindEye2 \cite{mindeye2} focuses on high-fidelity brain-to-image decoding under limited data conditions, proposing a unified backbone that maps multi-subject fMRI signals into a shared latent space before projecting them into image feature space. By combining contrastive objectives with reconstruction losses, MindEye2 achieves strong image reconstruction quality with minimal per-subject data, particularly in multi-subject scenarios. Similarly, VINDEX \cite{vindex} emphasizes recovering essential visual details from neural signals by introducing multi-granularity image features, enabling more precise neural decoding across different semantic levels.

Other approaches aim to strengthen neural-visual correspondence through representation learning techniques. MindBridge \cite{mindbridge} constructs a unified fMRI-image-text manifold trained with a cyclic consistency objective, encouraging bidirectional reconstruction between projected brain activity and CLIP-based semantic embeddings. MindArtist \cite{mindart} formulates alignment as an optimal transport problem with graph matching, decomposing neural representations into semantic content and contextual style to enable controllable neural-to-image generation. Contrastive learning has also been widely adopted: BrainCLIP \cite{brainclip} aligns fMRI embeddings with both CLIP visual and text features using a dual contrastive objective, while MindTuner \cite{mindtuner} introduces a visual pivot mechanism that anchors neural signals to dense visual features before bridging them to language representations.

While these methods differ in supervision and architecture, they are primarily motivated by \textbf{improving alignment quality or reconstruction fidelity}, often assuming access to \textbf{complete visual information}. In contrast, our MindVLM is designed for \textbf{brain captioning under partial visual observability}, where the goal is not to recover full image representations but to identify which visual concepts are supported by neural activity. Rather than enforcing point-wise matching between fMRI signals and global image embeddings, MindVLM derives global semantic representations from local visual tokens and uses them to perform semantic evidence selection. This formulation preserves token-level interpretability while avoiding over-constraining neural representations to visual details that may not be neurally observed, enabling more faithful and subject-aware brain captioning.

\subsubsection{Brain-to-LLM Integration}

With the rapid advancement of Large Language Models (LLMs), recent work on brain-to-text captioning has begun to directly inject fMRI signals into the token space of pretrained LLMs, enabling the model to generate captions or answers without relying on an explicit visual encoder. OneLLM \cite{onellm} initiates this direction by introducing Universal Projection Modules that map fMRI signals directly into the token embedding space of a frozen LLaMA2 \cite{llama}. This treats brain activity as a new ``language modality,'' progressively aligned through instruction tuning without requiring an intermediary visual representation. MindLLM \cite{mindllm} further advances this idea through Brain Instruction Tuning, in which fMRI features are encoded as prefix tokens to an LLM. This subject-agnostic encoder is optimized end-to-end, casting neural decoding as a pure language modeling task applicable to captioning and VQA. However, these approaches overlook a key issue that the text captions used as supervision are not naturally paired with the fMRI signals--the true pairing exists between \textbf{image and fMRI}. As a result, training with text-only losses risks drifting away from the actual perceptual semantics represented in the brain.

To address this, MindGPT \cite{mindgpt} reintroduces a visual grounding mechanism by using a visually guided neural encoder with cross-attention to map fMRI signals into the shared CLIP space. By distilling CLIP's joint vision-language semantics into the brain encoder, it allows the downstream GPT decoder to produce captions that are more faithful to the viewed images. A related framework proposed by \citep{nv2lan} employs a 3D ViT \cite{vit} to extract fMRI features and hierarchically align them with multi-level visual representations (low-level structure and high-level semantics) from a pretrained vision encoder before projecting them into the LLM input space. This enables ``brain-visual question answering,'' where neural activity functions as the model's visual context. More recently, MindSA \cite{mindsa} solves a different limitation that the inherent mismatch between entire images and the selective perceptual content that subjects actually attend to. By dynamically masking out visual regions using a differentiable semantic removal mechanism, it creates brain-compatible supervision for a frozen GPT-2 decoder and ensures that generated captions reflect only the neurally processed semantics.

Despite their architectural diversity, all the models above share a critical constraint: they require a frozen LLM during training, because the alignment or integration is optimized through text-based supervision. This increases computational cost and risks introducing biases from textual priors unrelated to the subject's perceptual state. UMBRAE \cite{umbrae} mitigates this by training a universal brain encoder that maps subject-specific fMRI signals into a shared space approximating the visual tokens of a multi-modal LLM. This ``Perceiver Encoder'' enables brain signals to serve directly as visual prompts, bypassing the need for explicit text decoding during training. Our \textbf{MindVLM} follows this elegance of removing heavy LLMs from the training loop. However, instead of forcing fMRI representations to exactly match frozen image/text features, which can over-constrain learning and promote ``over-imagining,'' we perform adaptive alignment that preserves flexibility while maintaining semantic fidelity. Moreover, by aligning at local-token derived global-level representations of visual and neural modalities, our \textbf{MindVLM} naturally supports interpretable neural-to-language captioning, enabling us to trace which visual tokens contribute to each part of the generated caption during inference.

\subsection{Dataset and Implementation}
\label{sec:data}

\subsubsection{NSD Dataset}
The Natural Scenes Dataset (NSD) \citep{nsd} is a large-scale neuroimaging benchmark designed to capture high-resolution fMRI responses during natural scene perception. It contains whole-brain 7T fMRI recordings from eight participants at 1.8 mm isotropic resolution, collected while subjects viewed a diverse set of real-world images. Each participant completed 30-40 scanning sessions and viewed 10,000 images (9,000 unique to each subject plus 1,000 shared across all subjects), for a total of 70,566 trials. All visual stimuli were selected from the Microsoft COCO (MS-COCO) \cite{mscoco} dataset \citep{mscoco}, a richly annotated corpus where each image is paired with five human-written captions describing its semantic content. However, to accommodate the MRI display system and experimental design, the original MS-COCO images--typically rectangular and much larger--were center-cropped and resized to a 425$\times$425 square resolution before being shown to participants. As a consequence, certain objects mentioned in the captions may not appear in the cropped NSD stimulus, creating a systematic semantic mismatch between the visual input that drives the neural responses and the full MS-COCO caption set.

During the neural experiment, each image was presented for 3 seconds, accompanied by an object-recognition task to ensure sustained attention and to probe recognition and memory processes. The NSD pre-processing pipeline provides high-quality single-trial beta estimates obtained through voxel-wise hemodynamic response modeling and advanced denoising, yielding exceptionally high single-trial signal-to-noise ratios. In addition to the functional recordings, NSD includes anatomical MRI, resting-state scans, and detailed behavioral performance measures. Functional data are delivered in both subject-native cortical surface space and standardized volumetric spaces such as fsaverage \citep{fsaverage} and MNI152 \citep{mni152}, enabling flexible integration with popular neuro-imaging tools including FreeSurfer \citep{freesurfer} and FSL \cite{fsl}. Furthermore, NSD provides manually annotated regions of interest (ROIs), including retinotopic visual areas and high-level category-selective regions along the ventral visual stream. In this paper, we adopt the \textit{nsdgeneral} atlas, which yields various number of voxels: 15,724 (\textit{subj01}), 14,278 (\textit{subj02}), 13,039 (\textit{subj05}), and 12,682 (\textit{subj07}).

With its scale, annotation richness, and high spatial fidelity, NSD has become a central resource for modeling human vision and advancing brain decoding methods that integrate neural, visual, and linguistic information.

\subsubsection{Implementation Details}
\label{subsec:impl}
In the main paper, we mainly report results using MindVLM with LLaVA \cite{llava} as the underlying vision-language model. Specifically, we adopt the Stage-1 LLaVA-v1.5-7B configuration, where both the language model and the visual encoder (OpenAI CLIP ViT-L/14-336) remain completely frozen throughout training. To facilitate efficient contrastive alignment, we precompute visual features for all images in the NSD dataset and load them directly during training. The fMRI encoder is a 4-layer 4-head Vision Transformer \cite{vit}, following the architecture design of \cite{mindvis}, and is trained from scratch for our alignment task. We train the model for 150 epochs, including 10 warm-up epochs, using a batch size of 1000. The learning rate is initialized at $7.5\times10^{-3}$ and decayed to zero using a cosine annealing schedule. Optimization is performed using AdamW with a weight decay of 0.05 and gradient clipping set to 0.8. All hyperparameters are kept consistent across subjects unless otherwise specified. Experiments are run on two NVIDIA RTX 6000 Ada GPUs. Training requires about 45 minutes per subject. During inference, we set the maximum number of new tokens the VLM can generate to 32 and use the system message as ``You are a helpful assistant.'' A fixed random seed of 2025 is set for all experiments in both training and evaluation.

\subsection{Detailed Captioning Performance By Subject}\label{subsec:cap_res_all}
Similar to Table 1 in the main paper, we provide a detailed captioning performance of each subject at all thresholds of $q\%$.

\begin{table}[ht]
\centering
\caption{NSD captioning evaluation for subj01 with MindVLM-LLaVA.}\vspace{-3mm}
\resizebox{\textwidth}{!}{%
\begin{tabular}{c|l|cccc|cccc}
\toprule
                                    &               & \multicolumn{4}{c|}{\textbf{BLEU-n Reference}} & \multicolumn{4}{c}{\textbf{Content \& Semantic Agreement}} \\
\cmidrule(lr){3-6} \cmidrule(lr){7-10}
\textbf{Subject}    & \textbf{Model} & \textbf{B@1} & \textbf{B@2} & \textbf{B@3} & \textbf{B@4} & \textbf{ROUGE} & \textbf{METEOR} & \textbf{CIDEr} & \textbf{SPICE} \\
\midrule
\multirow{20}{*}{subj01} 
& MindVLM ($q=1 $) & 44.33 & 30.40 & 19.93 & 12.98 & 39.05 & 23.42 & 21.28 & 17.12 \\
& MindVLM ($q=5 $) & 48.34 & 33.97 & 22.59 & 14.74 & 42.24 & 25.03 & 35.72 & 19.22 \\
& MindVLM ($q=10$) & 49.88 & 35.27 & 23.64 & 15.44 & 43.47 & 25.63 & 41.88 & 19.92 \\
& MindVLM ($q=15$) & 50.85 & 36.15 & 24.24 & 15.86 & 44.06 & 25.96 & 45.10 & 20.24 \\
& MindVLM ($q=20$) & 52.14 & 37.37 & 25.26 & 16.63 & 44.84 & 26.31 & 48.67 & 20.80 \\
& MindVLM ($q=25$) & 53.10 & 38.12 & 25.79 & 17.00 & 45.46 & 26.73 & 52.19 & 21.03 \\
& MindVLM ($q=30$) & 53.97 & 38.74 & 26.26 & 17.35 & 45.94 & 27.09 & 54.46 & 21.15 \\
& MindVLM ($q=35$) & 54.22 & 38.98 & 26.46 & 17.56 & 46.21 & 27.11 & 55.96 & 21.16 \\
& MindVLM ($q=40$) & 55.18 & 39.81 & 26.99 & 17.87 & 46.94 & 27.21 & 58.68 & 21.47 \\
& MindVLM ($q=45$) & 55.29 & 40.05 & 27.34 & 18.27 & 47.16 & 27.41 & 59.11 & 21.55 \\
& MindVLM ($q=50$) & 55.66 & 40.23 & 27.48 & \textbf{18.38} & 47.39 & \textbf{27.50} & 59.56 & 21.69 \\
& MindVLM ($q=55$) & 56.12 & 40.44 & 27.55 & 18.31 & \textbf{47.55} & 27.46 & 60.79 & 21.67 \\
& MindVLM ($q=60$) & 55.98 & 40.26 & 27.37 & 18.16 & 47.47 & 27.40 & 60.41 & \textbf{21.72} \\
& MindVLM ($q=65$) & 55.97 & 40.32 & 27.38 & 18.13 & 47.42 & 27.41 & 60.22 & 21.69 \\
& MindVLM ($q=70$) & \textbf{56.21} & \textbf{40.55} & \textbf{27.65} & 18.35 & 47.53 & 27.42 & \textbf{62.06} & 21.61 \\
& MindVLM ($q=75$) & 55.91 & 40.39 & 27.57 & 18.33 & 47.51 & 27.33 & 61.57 & 21.52 \\
& MindVLM ($q=80$) & 55.99 & 40.34 & 27.50 & 18.26 & 47.49 & 27.31 & 61.68 & 21.56 \\
& MindVLM ($q=85$) & 55.86 & 40.25 & 27.42 & 18.21 & 47.37 & 27.30 & 61.41 & 21.50 \\
& MindVLM ($q=90$) & 55.90 & 40.29 & 27.47 & 18.24 & 47.39 & 27.28 & 61.43 & 21.49 \\
& MindVLM ($q=95$) & 55.88 & 40.28 & 27.45 & 18.23 & 47.39 & 27.28 & 61.40 & 21.49 \\
\bottomrule
\end{tabular}}
\label{tb:cap_res_s1}
\end{table}

\begin{table}[t]
\centering
\caption{NSD captioning evaluation for subj02 with MindVLM-LLaVA.}\vspace{-3mm}
\resizebox{\textwidth}{!}{%
\begin{tabular}{c|l|cccc|cccc}
\toprule
                                    &               & \multicolumn{4}{c|}{\textbf{BLEU-n Reference}} & \multicolumn{4}{c}{\textbf{Content \& Semantic Agreement}} \\
\cmidrule(lr){3-6} \cmidrule(lr){7-10}
\textbf{Subject}    & \textbf{Model} & \textbf{B@1} & \textbf{B@2} & \textbf{B@3} & \textbf{B@4} & \textbf{ROUGE} & \textbf{METEOR} & \textbf{CIDEr} & \textbf{SPICE} \\
\midrule
\multirow{20}{*}{subj02} 
& MindVLM($q=1$) & 44.07 & 30.05 & 19.65 & 12.74 & 38.64 & 23.06 & 19.47 & 17.09 \\
& MindVLM($q=5$) & 47.24 & 32.98 & 21.84 & 14.21 & 41.04 & 24.29 & 30.16 & 18.28 \\
& MindVLM($q=10$) & 49.38 & 34.65 & 23.00 & 15.03 & 42.84 & 25.19 & 39.00 & 19.43 \\
& MindVLM($q=15$) & 50.50 & 35.86 & 23.89 & 15.59 & 43.59 & 25.72 & 42.99 & 19.93 \\
& MindVLM($q=20$) & 51.41 & 36.65 & 24.55 & 16.05 & 44.24 & 26.00 & 45.52 & 20.27 \\
& MindVLM($q=25$) & 51.98 & 37.21 & 24.92 & 16.32 & 44.67 & 26.24 & 48.30 & 20.64 \\
& MindVLM($q=30$) & 53.07 & 38.06 & 25.65 & 16.85 & 45.42 & 26.53 & 52.47 & 20.65 \\
& MindVLM($q=35$) & 54.18 & 38.97 & 26.44 & 17.49 & 46.08 & 26.87 & 55.27 & 20.99 \\
& MindVLM($q=40$) & 54.44 & 39.16 & 26.52 & 17.52 & 46.52 & 26.93 & 56.64 & 20.93 \\
& MindVLM($q=45$) & 55.03 & 39.59 & 26.88 & 17.77 & 46.74 & 27.20 & 57.95 & 21.24 \\
& MindVLM($q=50$) & 54.99 & 39.60 & 26.97 & 17.88 & 46.81 & 27.14 & 58.14 & 21.22 \\
& MindVLM($q=55$) & 55.75 & 40.36 & 27.70 & 18.47 & 47.44 & \textbf{27.40} & 60.23 & \textbf{21.54} \\
& MindVLM($q=60$) & 55.81 & 40.42 & 27.68 & 18.42 & 47.50 & 27.36 & 59.97 & 21.40 \\
& MindVLM($q=65$) & 55.48 & 40.09 & 27.34 & 18.22 & 47.34 & 27.34 & 60.58 & 21.38 \\
& MindVLM($q=70$) & 55.86 & 40.47 & 27.72 & 18.50 & 47.65 & 27.38 & 61.64 & 21.52 \\
& MindVLM($q=75$) & 55.79 & 40.38 & 27.65 & 18.45 & 47.60 & 27.34 & 61.53 & 21.47 \\
& MindVLM($q=80$) & 56.06 & 40.55 & 27.73 & 18.50 & 47.62 & 27.29 & 62.38 & 21.46 \\
& MindVLM($q=85$) & 56.10 & 40.51 & 27.70 & 18.48 & 47.70 & 27.33 & 63.08 & 21.49 \\
& MindVLM($q=90$) & \textbf{56.15} & \textbf{40.57} & \textbf{27.77} & \textbf{18.57} & \textbf{47.73} & 27.32 & \textbf{63.28} & 21.49 \\
& MindVLM($q=95$) & 56.12 & 40.55 & 27.73 & 18.53 & 47.71 & 27.32 & 63.15 & 21.48 \\
\bottomrule
\end{tabular}}
\label{tb:cap_res_s2}
\end{table}

\begin{table}[t]
\centering
\caption{NSD captioning evaluation for subj05 with MindVLM-LLaVA.}\vspace{-3mm}
\resizebox{\textwidth}{!}{%
\begin{tabular}{c|l|cccc|cccc}
\toprule
                                    &               & \multicolumn{4}{c|}{\textbf{BLEU-n Reference}} & \multicolumn{4}{c}{\textbf{Content \& Semantic Agreement}} \\
\cmidrule(lr){3-6} \cmidrule(lr){7-10}
\textbf{Subject}    & \textbf{Model} & \textbf{B@1} & \textbf{B@2} & \textbf{B@3} & \textbf{B@4} & \textbf{ROUGE} & \textbf{METEOR} & \textbf{CIDEr} & \textbf{SPICE} \\
\midrule
\multirow{20}{*}{subj05} 
& MindVLM($q=1$) & 43.59 & 29.65 & 19.29 & 12.55 & 38.22 & 23.05 & 16.94 & 16.73 \\
& MindVLM($q=5$) & 47.10 & 32.96 & 21.69 & 14.08 & 41.17 & 24.50 & 31.17 & 18.51 \\
& MindVLM($q=10$) & 49.14 & 34.67 & 23.04 & 14.96 & 42.56 & 25.22 & 37.73 & 19.57 \\
& MindVLM($q=15$) & 49.92 & 35.33 & 23.56 & 15.38 & 43.24 & 25.67 & 41.11 & 19.87 \\
& MindVLM($q=20$) & 50.54 & 35.81 & 23.96 & 15.73 & 43.88 & 25.95 & 43.06 & 20.09 \\
& MindVLM($q=25$) & 51.97 & 36.97 & 24.76 & 16.27 & 44.75 & 26.38 & 48.21 & 20.51 \\
& MindVLM($q=30$) & 52.76 & 37.70 & 25.39 & 16.74 & 45.45 & 26.53 & 51.88 & 20.93 \\
& MindVLM($q=35$) & 53.57 & 38.41 & 25.83 & 17.05 & 45.92 & 26.80 & 53.70 & 21.17 \\
& MindVLM($q=40$) & 54.21 & 38.99 & 26.42 & 17.51 & 46.31 & 27.03 & 55.18 & 21.41 \\
& MindVLM($q=45$) & 54.90 & 39.57 & 26.94 & 17.97 & 46.85 & 27.22 & 56.64 & 21.47 \\
& MindVLM($q=50$) & 55.45 & 40.00 & 27.20 & 18.09 & 47.23 & 27.17 & 59.01 & 21.55 \\
& MindVLM($q=55$) & 55.16 & 39.76 & 27.05 & 17.99 & 47.11 & 27.25 & 58.92 & \textbf{21.66} \\
& MindVLM($q=60$) & 55.60 & 40.13 & 27.30 & 18.17 & 47.29 & 27.39 & 59.85 & 21.51 \\
& MindVLM($q=65$) & 55.59 & 39.91 & 27.12 & 18.08 & 47.36 & 27.39 & 59.65 & 21.56 \\
& MindVLM($q=70$) & 55.68 & 40.11 & 27.32 & \textbf{18.26} & 47.38 & \textbf{27.43} & 59.51 & 21.50 \\
& MindVLM($q=75$) & 55.78 & \textbf{40.16} & \textbf{27.34} & 18.20 & 47.38 & 27.43 & 60.46 & 21.43 \\
& MindVLM($q=80$) & 55.76 & 40.15 & 27.32 & 18.13 & 47.44 & 27.41 & 61.03 & 21.50 \\
& MindVLM($q=85$) & 55.67 & 40.06 & 27.28 & 18.08 & 47.50 & 27.42 & 60.70 & 21.58 \\
& MindVLM($q=90$) & 55.79 & 40.07 & 27.25 & 18.06 & 47.49 & 27.41 & 60.78 & 21.53 \\
& MindVLM($q=95$) & \textbf{55.85} & 40.14 & 27.31 & 18.10 & \textbf{47.58} & 27.43 & \textbf{61.06} & 21.56 \\
\bottomrule
\end{tabular}}
\label{tb:cap_res_s5}
\end{table}

\begin{table}[t]
\centering
\caption{NSD captioning evaluation for subj07 with MindVLM-LLaVA.}\vspace{-3mm}
\resizebox{\textwidth}{!}{%
\begin{tabular}{c|l|cccc|cccc}
\toprule
                                    &               & \multicolumn{4}{c|}{\textbf{BLEU-n Reference}} & \multicolumn{4}{c}{\textbf{Content \& Semantic Agreement}} \\
\cmidrule(lr){3-6} \cmidrule(lr){7-10}
\textbf{Subject}    & \textbf{Model} & \textbf{B@1} & \textbf{B@2} & \textbf{B@3} & \textbf{B@4} & \textbf{ROUGE} & \textbf{METEOR} & \textbf{CIDEr} & \textbf{SPICE} \\
\midrule
\multirow{20}{*}{subj07} 
& MindVLM($q=1$) & 44.18 & 30.15 & 19.78 & 12.93 & 38.94 & 23.22 & 19.50 & 17.16 \\
& MindVLM($q=5$) & 47.73 & 33.48 & 22.20 & 14.54 & 41.55 & 24.74 & 33.24 & 18.83 \\
& MindVLM($q=10$) & 49.49 & 34.84 & 23.45 & 15.54 & 43.14 & 25.48 & 41.22 & 19.45 \\
& MindVLM($q=15$) & 50.74 & 35.88 & 24.00 & 15.68 & 43.86 & 25.87 & 45.98 & 19.82 \\
& MindVLM($q=20$) & 52.33 & 37.22 & 25.13 & 16.50 & 44.91 & 26.26 & 51.14 & 20.14 \\
& MindVLM($q=25$) & 52.71 & 37.53 & 25.26 & 16.53 & 45.16 & 26.40 & 51.27 & 20.45 \\
& MindVLM($q=30$) & 53.49 & 38.34 & 25.94 & 17.12 & 45.50 & 26.71 & 53.18 & 20.50 \\
& MindVLM($q=35$) & 54.17 & 38.83 & 26.12 & 17.17 & 45.81 & 26.81 & 54.35 & 20.84 \\
& MindVLM($q=40$) & 54.72 & 39.22 & 26.44 & 17.33 & 46.25 & 26.96 & 56.63 & 21.00 \\
& MindVLM($q=45$) & 54.87 & 39.32 & 26.58 & 17.60 & 46.46 & 26.96 & 58.46 & 21.06 \\
& MindVLM($q=50$) & 55.78 & 39.94 & 27.04 & 17.89 & 46.95 & 27.12 & 61.37 & 21.53 \\
& MindVLM($q=55$) & 56.06 & 40.31 & 27.49 & 18.30 & 47.35 & 27.32 & 61.91 & 21.59 \\
& MindVLM($q=60$) & 55.77 & 40.06 & 27.20 & 18.04 & 47.38 & 27.25 & 62.12 & 21.36 \\
& MindVLM($q=65$) & 56.19 & 40.34 & 27.44 & 18.22 & 47.58 & 27.38 & 62.61 & 21.41 \\
& MindVLM($q=70$) & 56.51 & 40.58 & 27.51 & 18.24 & 47.68 & \textbf{27.39} & 62.98 & \textbf{21.70} \\
& MindVLM($q=75$) & 56.52 & 40.57 & 27.48 & 18.23 & \textbf{47.81} & 27.37 & 62.61 & 21.64 \\
& MindVLM($q=80$) & 56.43 & 40.52 & 27.44 & 18.17 & 47.75 & 27.38 & 62.79 & 21.52 \\
& MindVLM($q=85$) & 56.55 & 40.69 & 27.62 & 18.32 & 47.77 & 27.35 & 63.36 & 21.47 \\
& MindVLM($q=90$) & 56.61 & 40.74 & \textbf{27.66} & 18.35 & 47.81 & 27.37 & 63.55 & 21.49 \\
& MindVLM($q=95$) & \textbf{56.62} & \textbf{40.75} & 27.66 & \textbf{18.36} & 47.80 & 27.37 & \textbf{63.58} & 21.48 \\
\bottomrule
\end{tabular}}
\label{tb:cap_res_s7}
\end{table}

\begin{table}[t]
\centering
\caption{NSD captioning evaluation average over four subjects with MindVLM-LLaVA.}\vspace{-3mm}
\resizebox{\textwidth}{!}{%
\begin{tabular}{c|l|cccc|cccc}
\toprule
                                    &               & \multicolumn{4}{c|}{\textbf{BLEU-n Reference}} & \multicolumn{4}{c}{\textbf{Content \& Semantic Agreement}} \\
\cmidrule(lr){3-6} \cmidrule(lr){7-10}
\textbf{Subject}    & \textbf{Model} & \textbf{B@1} & \textbf{B@2} & \textbf{B@3} & \textbf{B@4} & \textbf{ROUGE} & \textbf{METEOR} & \textbf{CIDEr} & \textbf{SPICE} \\
\midrule
\multirow{20}{*}{average} 
& MindVLM($q=1$) & 44.04 & 30.06 & 19.66 & 12.80 & 38.71 & 23.19 & 19.30 & 17.03  \\
& MindVLM($q=5$) & 47.60 & 33.34 & 22.08 & 14.39 & 41.50 & 24.64 & 32.57 & 18.71  \\
& MindVLM($q=10$) & 49.47 & 34.86 & 23.28 & 15.24 & 43.00 & 25.38 & 39.96 & 19.59 \\
& MindVLM($q=15$) & 50.50 & 35.81 & 23.92 & 15.63 & 43.69 & 25.80 & 43.80 & 19.97 \\
& MindVLM($q=20$) & 51.61 & 36.76 & 24.73 & 16.23 & 44.47 & 26.13 & 47.10 & 20.32 \\
& MindVLM($q=25$) & 52.44 & 37.46 & 25.18 & 16.53 & 45.01 & 26.44 & 49.99 & 20.66 \\
& MindVLM($q=30$) & 53.33 & 38.21 & 25.81 & 17.01 & 45.58 & 26.71 & 53.00 & 20.81 \\
& MindVLM($q=35$) & 54.03 & 38.80 & 26.21 & 17.32 & 46.00 & 26.90 & 54.82 & 21.04 \\
& MindVLM($q=40$) & 54.64 & 39.30 & 26.59 & 17.56 & 46.50 & 27.03 & 56.78 & 21.20 \\
& MindVLM($q=45$) & 55.03 & 39.63 & 26.93 & 17.90 & 46.80 & 27.20 & 58.04 & 21.33 \\
& MindVLM($q=50$) & 55.47 & 39.94 & 27.17 & 18.06 & 47.09 & 27.23 & 59.52 & 21.49 \\
& MindVLM($q=55$) & 55.77 & 40.22 & 27.45 & 18.27 & 47.36 & 27.36 & 60.46 & \textbf{21.61} \\
& MindVLM($q=60$) & 55.79 & 40.22 & 27.39 & 18.20 & 47.41 & 27.35 & 60.59 & 21.50 \\
& MindVLM($q=65$) & 55.81 & 40.16 & 27.32 & 18.16 & 47.43 & 27.38 & 60.77 & 21.51 \\
& MindVLM($q=70$) & 56.06 & 40.43 & \textbf{27.55} & \textbf{18.34} & 47.56 & \textbf{27.40} & 61.55 & 21.58 \\
& MindVLM($q=75$) & 56.00 & 40.37 & 27.51 & 18.30 & 47.57 & 27.37 & 61.54 & 21.51 \\
& MindVLM($q=80$) & 56.06 & 40.39 & 27.50 & 18.26 & 47.58 & 27.35 & 61.97 & 21.51 \\
& MindVLM($q=85$) & 56.05 & 40.38 & 27.51 & 18.27 & 47.59 & 27.35 & 62.14 & 21.51 \\
& MindVLM($q=90$) & 56.11 & 40.42 & 27.54 & 18.31 & 47.61 & 27.34 & 62.26 & 21.50 \\
& MindVLM($q=95$) & \textbf{56.12} & \textbf{40.43} & 27.54 & 18.30 & \textbf{47.62} & 27.35 & \textbf{62.30} & 21.50 \\
\bottomrule
\end{tabular}}
\label{tb:cap_res_avg}
\end{table}
\clearpage

\subsection{Visual Token Selection} 
\label{subsec:q_analysis_app}
\begin{figure}[t]
  \centering
  \includegraphics[width=\columnwidth]{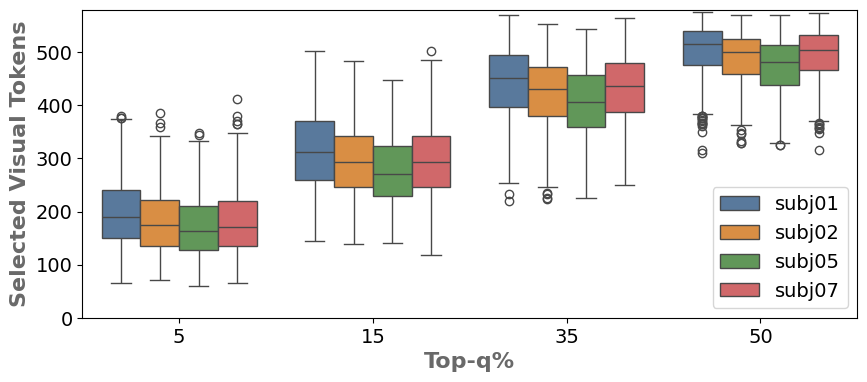}
  \vspace{-6mm}\caption{Number of selected visual tokens \emph{v.s.} threshold $q$.}\label{fig:q_ntokens} \vspace{-2mm}
\end{figure}
The hyperparameter $q$ controls the proportion of high-similarity neural-visual pairs retained (evidence) during neural-visual matching and thus determines how many visual tokens are considered attended by a subject. Figure \ref{fig:q_ntokens} visualizes the distribution of the number of selected tokens across four subjects for different $q$ thresholds (5, 15, 35, and 50). As expected, the number of selected visual tokens increases monotonically with $q$. At very low thresholds (e.g., $q=5$), only a small subset of highly confident neural-visual correspondences is preserved, resulting in a compact yet precise attentional representation (typically 150-250 tokens). As $q$ increases, more visual regions are included, gradually expanding the perceptual coverage but also introducing greater variability across samples. This trade-off between sparsity and completeness reflects how strict attention constraints (small $q$) emphasize the most salient perceptual features, whereas larger $q$ captures a broader, more diffuse perceptual field. Across subjects, the token distributions are broadly consistent, indicating the robustness of our neural-visual alignment mechanism. However, slight inter-subject variability can be observed--some subjects (e.g., subj01) tend to yield higher token counts, suggesting individual differences in attentional spread or representational granularity. These results confirm that $q$ acts as a controllable knob for tuning the balance between interpretability and perceptual coverage in subject-specific attention inference.
\clearpage

\subsection{Brain-Vision Matching}

The strong brain captioning performance in the previous section can be largely attributed to the effectiveness of our proposed contrastive matching between brain and visual representations. In this subsection, we turn our attention to directly evaluating this neural-visual mapping.

\begin{figure}[t]
  \centering
  \includegraphics[width=\columnwidth]{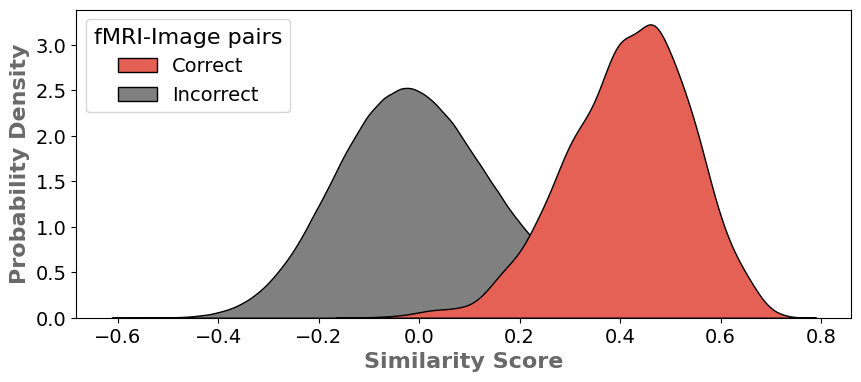}
  \vspace{-6mm}\caption{Similarity distribution of global fMRI-Image alignment.}\label{fig:nv_dist} \vspace{-2mm}
\end{figure}

\begin{table}[t]
  \caption{Evaluation for matching between fMRI and Image with MindVLM-LLaVA.}\vspace{-1mm}
  \label{tb:nv_align}
  \centering
  \begin{tabular}{c|c
      >{\columncolor{gray!5}}c
      >{\columncolor{red!3}}c
      >{\columncolor{green!4}}c}
    \toprule
    \textbf{Task} & \textbf{N-way} &
    \textbf{R@1} & \textbf{R@5} & \textbf{R@10} \\
    \midrule
    \multirow{4}{*}{fMRI-to-Image} & 1000 & 17.83 & 44.48 & 58.18 \\
                                   & 500  & 25.70 & 57.18 & 71.28 \\
                                   & 200  & 39.08 & 74.35 & 86.65 \\
                                   & 100  & 50.30 & 85.54 & 93.78 \\
    \midrule
    \multirow{4}{*}{Image-to-fMRI} & 1000 & 19.63 & 45.90 & 59.88 \\
                                   & 500  & 27.48 & 58.43 & 72.15 \\
                                   & 200  & 40.65 & 74.60 & 87.60 \\
                                   & 100  & 51.50 & 86.85 & 94.53 \\
    \bottomrule
  \end{tabular}\vspace{-4mm}
\end{table}

\vspace{1mm}\noindent\textbf{Distribution Analysis.} To quantify the learned correspondence, we analyze the distribution of similarity scores between fMRI and image embeddings across the \textbf{1,000 test samples}. This results in approximately one million fMRI-image pairs (including symmetric duplicates). We categorize these pairs into two groups: \textbf{correct pairs} (the true image-fMRI pairs, 1,000 in total, shown in red) and \textbf{incorrect pairs} (mismatched combinations, about 500k in total, shown in gray). According to Figure \ref{fig:nv_dist}, the similarity distribution reveals a clear separation between the two groups. The incorrect pairs exhibit near-zero mean similarity, whereas the correct pairs show a significantly higher mean similarity (greater than 0.4). This distinct margin indicates that the learned embedding space effectively captures meaningful cross-modal correspondences between brain activity and visual representations at a semantic level.

\vspace{1mm}\noindent\textbf{Retrieval Analysis.} We further assess the learned neural-visual matching through N-way retrieval tasks, a standard evaluation for cross-modal contrastive learning. As reported in Table \ref{tb:nv_align}, we conduct retrieval under varying candidate pool sizes of 100, 200, 500, and 1,000, and measure performance using Recall@K (R@1, R@5, and R@10). Both fMRI-to-image and image-to-fMRI retrieval directions are evaluated. Across all settings, MindVLM achieves consistently strong results, showing high retrieval accuracy even under large candidate pools. These results confirm that our model learns a discriminative and semantically coherent shared space, where neural activity and visual representations can be accurately matched in a bidirectional manner.



\clearpage
\subsection{Semantic Loss \& Decoding Consistency}
\label{subsec:loss}
\begin{figure}[t]
  \centering
  \includegraphics[width=\columnwidth]{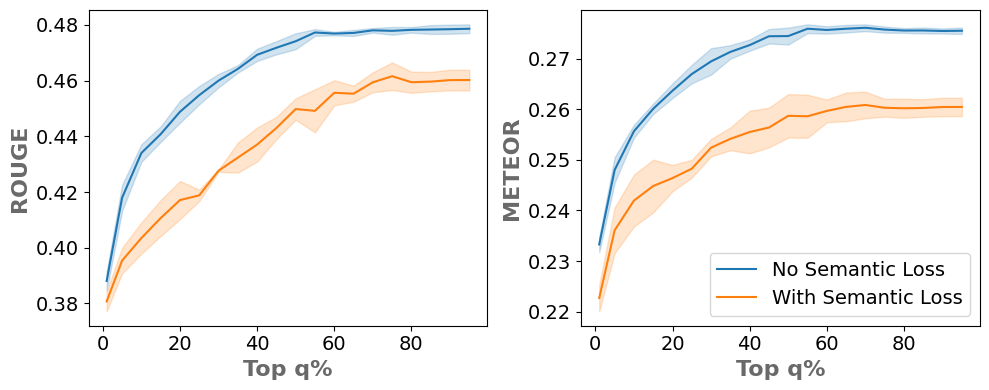}\vspace{-2mm}
  \caption{Performance comparison between two test sets with and without semantic loss, where we observe that our model shows a big difference in terms of two metrics ROUGE and METEOR.}\label{fig:metric_q_loss} \vspace{-5mm}
\end{figure}

To further examine the role of semantic selection in our paradigm, we compare performance between fMRI-image pairs with and without semantic loss, across varying $q$ thresholds. We report two representative metrics--ROUGE (n-gram based) and METEOR (semantic-based)--and include the rest in the Appendix \ref{subsec:loss_full}. As shown in Figure \ref{fig:metric_q_loss}, the curves reveal a consistent trend: samples without semantic loss outperform those with semantic loss across all $q$ levels, indicating the robustness of text-supervised fMRI-to-text decoding under well-aligned neural-visual-text triplets.

Quantitatively, the performance gap remains stable across thresholds: the no-semantic-loss group achieves approximately 0.47 ROUGE and 0.28 SPICE at $q=95$, compared to around 0.45 ROUGE and 0.26 SPICE for the semantic-loss group. The monotonic improvement of both curves with $q$ suggests that while larger token pools improve decoding fidelity, semantic consistency in supervision remains the dominant factor shaping model quality.


Overall, this result reinforces our central argument: accurate text supervision and semantically aligned fMRI-image pairs are indispensable for reliable fMRI-to-text decoding. Rather than relying on massive but noisy alignment sets, our paradigm advocates for selective and semantically consistent training--a shift that bridges neural representation learning and grounded language modeling.
\clearpage

\subsection{Data Leakage from MS-COCO into the VLM}
\label{app:leakage}
 
NSD stimuli come from MS-COCO, and VLMs can rely on memorized knowledge when answering questions~\cite{vlmbias}. We cannot rule out leakage by construction: LLaVA-v1.5's visual instruction tuning is built on MS-COCO images, and CLIP's pretraining corpus is not public. We therefore measure it for LLaVA-v1.5. Matching NSD's \texttt{cocoId} field against LLaVA's public instruction-tuning list (\texttt{llava\_v1\_5\_mix665k.json}) shows that LLaVA saw 232 of the 1,000 test images during instruction tuning (17.8\% across all 73,000 NSD images). We did not run this check for Qwen2.5-VL or Gemma3.
 
Table~\ref{tbl:leakage} splits the test set on this label and re-evaluates MindVLM-LLaVA-v1.5-7B at $q=35$, averaged over subjects, as in Table~3. BLEU here is the mean of per-image scores, which permits confidence intervals. It differs from corpus BLEU, which aggregates n-gram counts under a single brevity penalty and cannot be decomposed into subsets, so absolute values differ from Tables~1 and~8. The seen/unseen comparison is computed identically for both models.
 
\begin{table}[ht]
  \caption{Test images split by whether LLaVA saw them during instruction tuning (MindVLM-LLaVA at $q=35$, averaged over subjects). $\Delta$ is seen minus unseen; bold marks the higher score.}
  \label{tbl:leakage}
  \centering
  \begin{tabular}{llccc}
    \toprule
    Metric & Model & Seen (232) & Unseen (768) & $\Delta$ [95\% CI] \\
    \midrule
    BLEU@1 & LLaVA   & 58.52 & \textbf{60.36} & $-1.84$ [$-4.30$, $+0.63$] \\
           & MindVLM & 54.67 & \textbf{58.88} & $-4.22$ [$-6.16$, $-2.19$] \\
    \addlinespace
    BLEU@4 & LLaVA   & 11.74 & \textbf{13.19} & $-1.45$ [$-3.94$, $+1.17$] \\
           & MindVLM & 10.91 & \textbf{12.54} & $-1.63$ [$-3.64$, $+0.56$] \\
    \addlinespace
    ROUGE  & LLaVA   & 45.42 & \textbf{48.00} & $-2.58$ [$-4.46$, $-0.61$] \\
           & MindVLM & 43.08 & \textbf{46.89} & $-3.81$ [$-5.44$, $-2.12$] \\
    \addlinespace
    CIDEr  & LLaVA   & 49.27 & \textbf{64.21} & $-14.94$ [$-23.21$, $-6.15$] \\
           & MindVLM & 38.83 & \textbf{59.65} & $-20.82$ [$-27.53$, $-13.85$] \\
    \bottomrule
  \end{tabular}
\end{table}
 
Both models score lower on images LLaVA has seen, which runs counter to what memorization predicts. A reproduced caption would also appear in BLEU@4, which requires exact four-gram matches, yet neither model shows a significant seen/unseen difference on that metric. We therefore find no evidence that memorized COCO content improves captioning or differentially advantages MindVLM on NSD. We draw only this negative conclusion and do not interpret the size or direction of the gap: LLaVA's instruction-tuning subset was drawn from richly annotated, multi-object scenes, which are harder to caption and more likely to lose objects under NSD's crop, so the seen split is plausibly the harder one for reasons unrelated to exposure.

\subsection{Selected Tokens and Recorded Gaze}
\label{app:gaze}
To check whether the selected tokens fall on the parts of the stimulus that subjects looked at, we compare the masks selected by MindVLM-LLaVA at $q=5$, the sparse setting used for the subject-specific maps in Section~\ref{subsec:ssperception}, with NSD's eye-tracking records~\cite{nsd}. Because NSD subjects were instructed to maintain central fixation and generally complied, the recorded gaze reflects task-enforced fixation rather than free viewing, and the subject-specific differences we report reflect covert attention rather than where subjects looked. We therefore treat this comparison as a check of spatial concentration, not as a measure of attention.
 
For each test image, we define the fixated region as the visual tokens whose patches contain recorded fixation samples and compute its recall, the percentage of fixated-region tokens that selection retains. A random mask of the same size would, in expectation, recall the same percentage of the fixated region as the percentage of tokens it retains, which we use as size-matched chance.
 
\begin{table}[h]
  \caption{Tokens selected by MindVLM-LLaVA at $q=5$ against NSD's recorded gaze. Size-matched chance is the percentage of tokens retained.}
  \label{tbl:gaze}
  \centering
  \begin{tabular}{lcccc}
    \toprule
    & subj01 & subj02 & subj05 & subj07 \\
    \midrule
    Tokens retained (\%, size-matched chance) & 34.34 & 31.39 & 29.79 & 31.10 \\
    Fixated-region recall (\%) & 77.57 & 74.84 & 72.66 & 78.75 \\
    Ratio to chance & 2.26$\times$ & 2.38$\times$ & 2.44$\times$ & 2.53$\times$ \\
    \bottomrule
  \end{tabular}
\end{table}
 
As Table~\ref{tbl:gaze} shows, selection recovers the fixated region at 2.26-2.53 times size-matched chance, consistently across all four subjects. Because fixation was central, this comparison cannot separate subject-specific attention from a shared bias toward the image center; testing the former directly would require a free-viewing dataset.
 
\subsection{Per-Voxel Selection versus a Pooled fMRI Embedding}
\label{app:pooling}
Selection could instead use a single fMRI embedding pooled over voxel tokens. Because the projection in Eq.~(8) is linear, the similarity between the mean-pooled fMRI embedding and each visual token equals the mean of the corresponding column of $\mathbf{A}$ in Eq.~(9). Pooling is therefore a special case of our formulation: both read the same matrix, but pooling ranks a visual token by its mean similarity over voxel tokens, whereas the top-$q\%$ rule retains a token if any of its voxel pairs rank highly.
 
For MindVLM-LLaVA and each test image, we build a pooled mask of the same size as the per-voxel mask by retaining the visual tokens with the highest column means of $\mathbf{A}$. Table~\ref{tbl:pooling} reports the agreement between the two masks, the percentage of selected tokens that both masks keep, and the Gini coefficient of voxel participation, that is, of the number of selected pairs each voxel token contributes, which is 0 when all voxel tokens contribute equally.
 
\begin{table}[h]
  \caption{Per-voxel selection against a pooled fMRI embedding with the same mask size. Ranges are across the four subjects using MindVLM-LLaVA.}
  \label{tbl:pooling}
  \centering
  \begin{tabular}{ccc}
    \toprule
    $q$ & Agreement with pooled mask (\%) & Gini of voxel participation \\
    \midrule
    5  & $\sim$88 & 0.38-0.40 \\
    35 & $\sim$96 & 0.11-0.12 \\
    \bottomrule
  \end{tabular}
\end{table}
 
At $q=35$, voxel participation is close to uniform, and the two masks nearly coincide. At $q=5$, a few voxel tokens contribute most of the selected pairs, and the masks diverge. Since agreement rises with $q$, the setting of Table~\ref{tbl:vlm_compat} ($q=70$) does not depend on per-voxel modeling: pooling would recover nearly the same mask. The two formulations separate under sparse selection, where the subject-specific maps of Section~\ref{subsec:ssperception} and Appendix~\ref{subsec:vis_more} are computed. We therefore present per-voxel selection as suited to sparse selection rather than as necessary in all settings.

\subsection{More Results on Visualization}
\label{sec:vis}
In this section, we present visualization figures that are the same as those in the main paper but at a finer subject-level granularity.

\subsubsection{Caption performance v.s. $q$ for  all subjects}\label{subsec:metric_q}
In Figure \ref{fig:metric_q} of the main paper, we show how the scaled BLEU-3 and SPICE metrics--averaged across subjects--vary with the visual-token ratio as the threshold $q\%$ increases. Here, we extend this analysis by visualizing the results separately for each subject and by including all six remaining captioning metrics.

\begin{figure*}[ht]
  \centering
  \includegraphics[width=\textwidth]{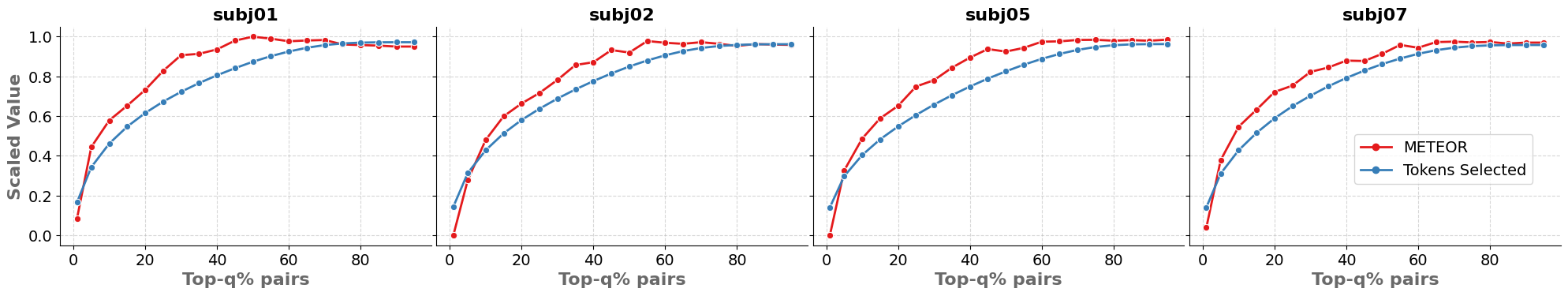}
  \includegraphics[width=\textwidth]{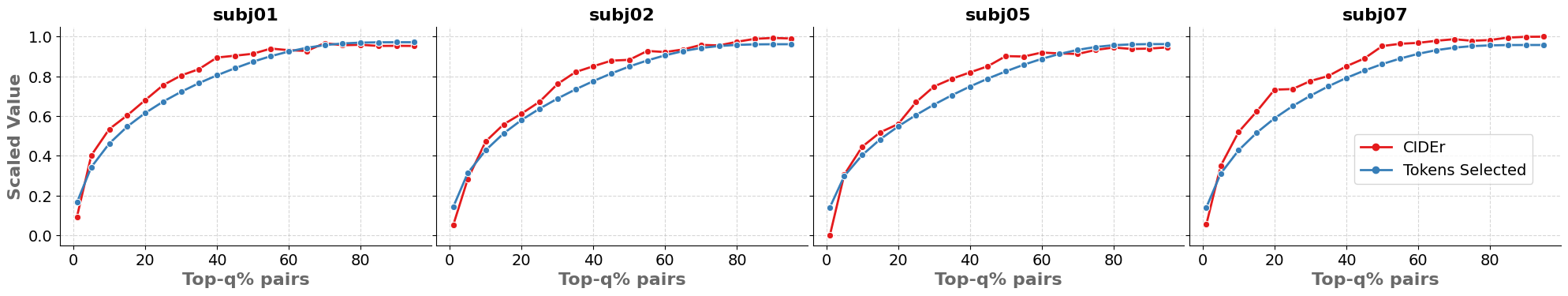}
  \includegraphics[width=\textwidth]{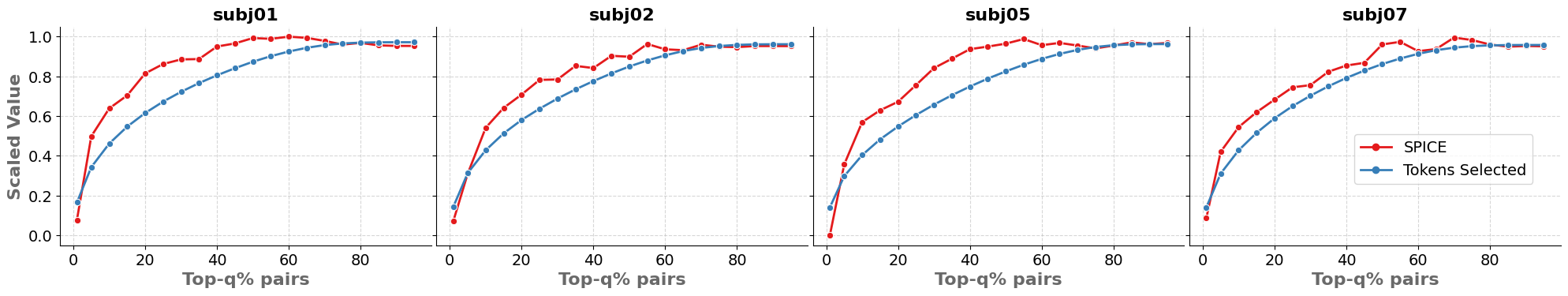}
  \caption{Scaled semantic metrics (METEOR, CIDEr, and SPICE) with visual tokens ratio \emph{v.s.} top $q\%$ selected fMRI-image token pairs.}\label{fig:q_cap_semantic}
\end{figure*}

\begin{figure*}[t]
  \centering
  \includegraphics[width=\textwidth]{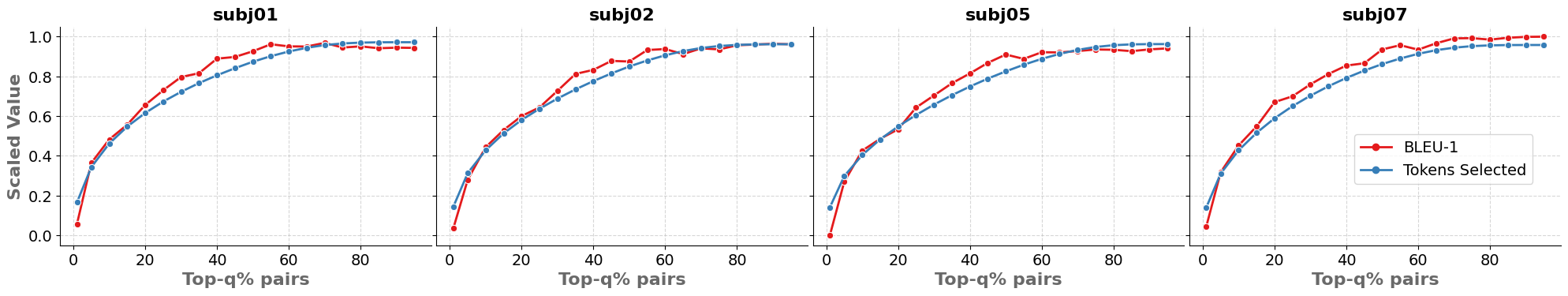}
  \includegraphics[width=\textwidth]{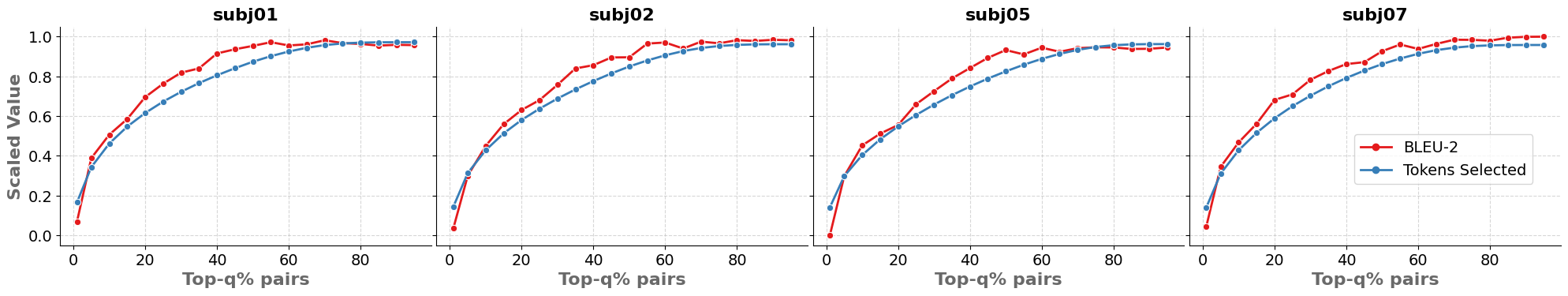}
  \includegraphics[width=\textwidth]{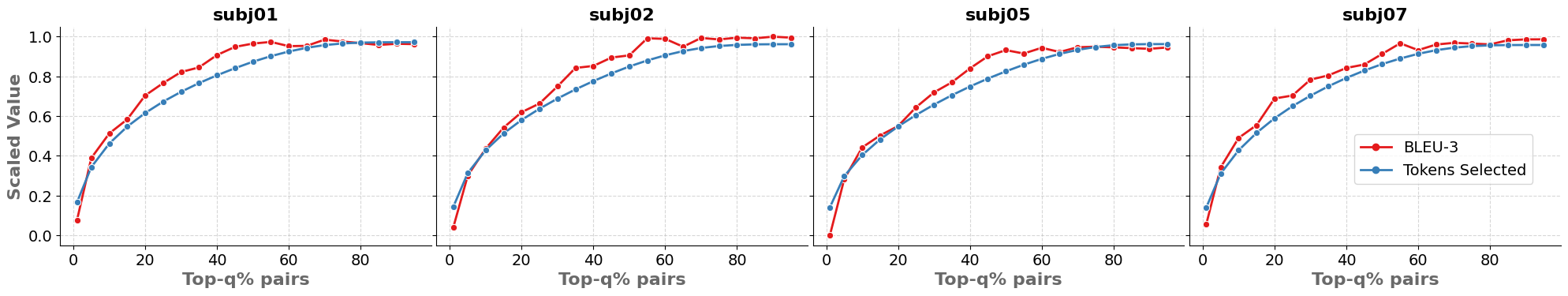}
  \includegraphics[width=\textwidth]{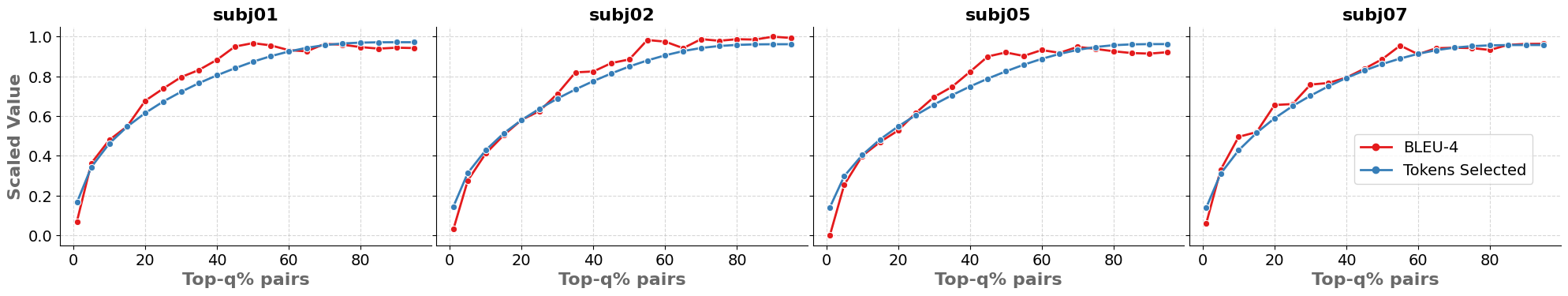}
  \includegraphics[width=\textwidth]{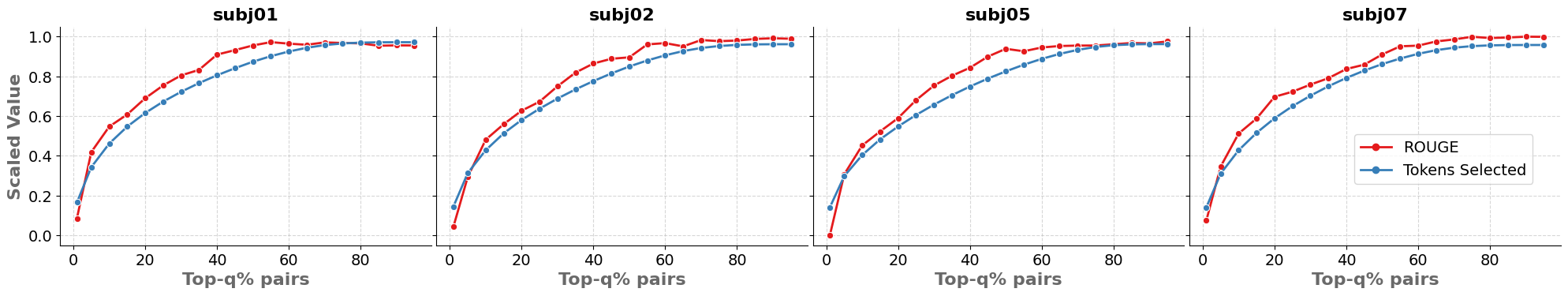}
  \vspace{-4mm}\caption{Scaled n-gram metrics (BLEU-1, BLEU-2, BLEU-3 ,BLEU-4, and ROUGE) with visual tokens ratio \emph{v.s.} top $q\%$ selected fMRI-image token pairs.}\label{fig:q_cap_ngram} \vspace{-5mm}
\end{figure*}

\clearpage

\subsubsection{Captions performance on samples with/without semantic loss}
\label{subsec:loss_full}
In Figure \ref{fig:metric_q_loss}, we show how the ROUGE and METEOR metrics--averaged across subjects--vary as the threshold $q\%$ increases for samples with and without semantic loss. Here, we extend this analysis by visualizing the results separately for each subject and by including all six remaining captioning metrics.

\begin{figure*}[ht]
  \centering
  \includegraphics[width=\textwidth]{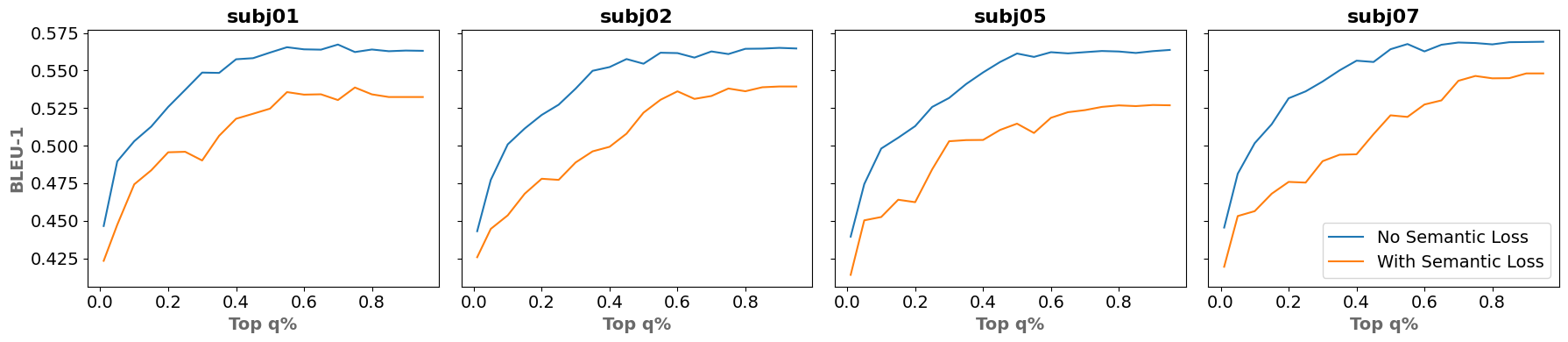}
  \includegraphics[width=\textwidth]{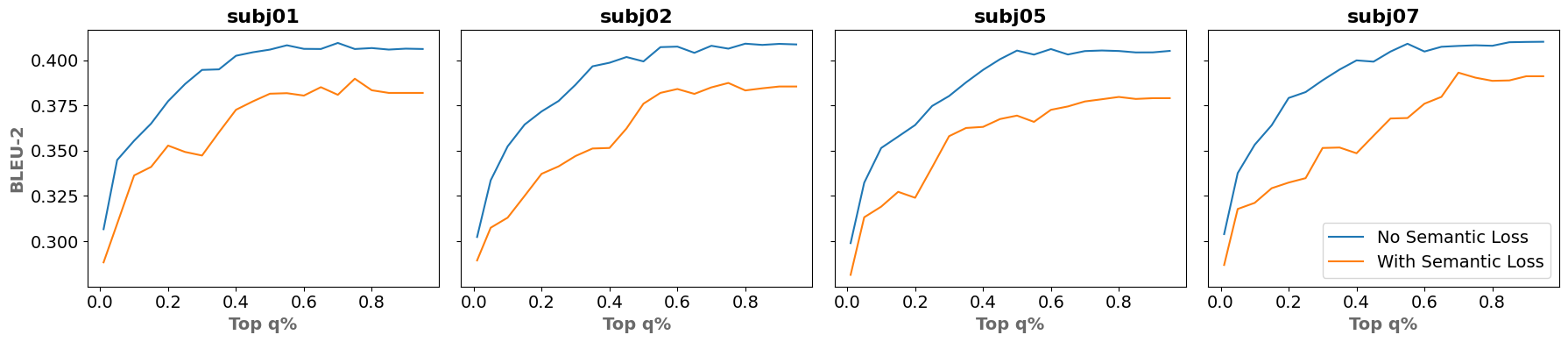}
  \includegraphics[width=\textwidth]{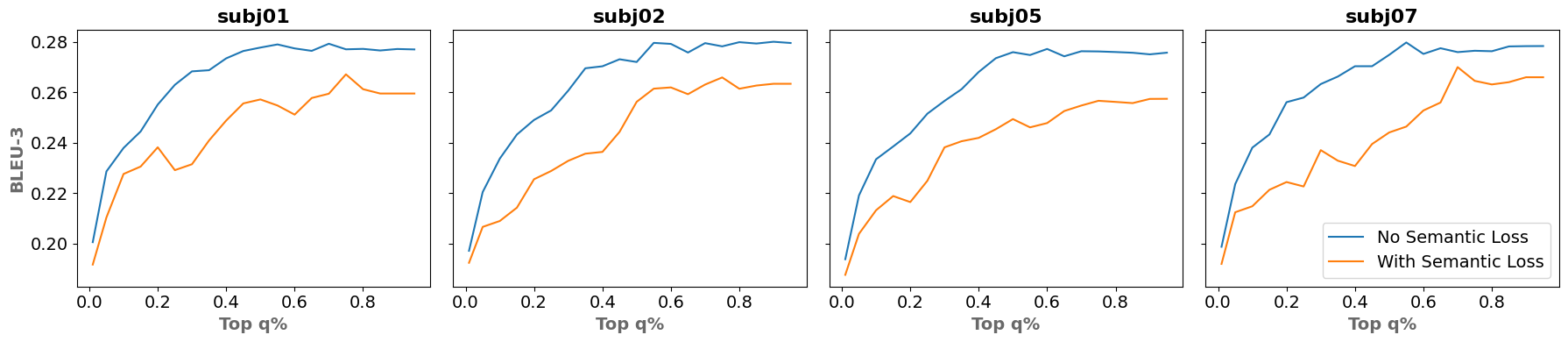}
  \includegraphics[width=\textwidth]{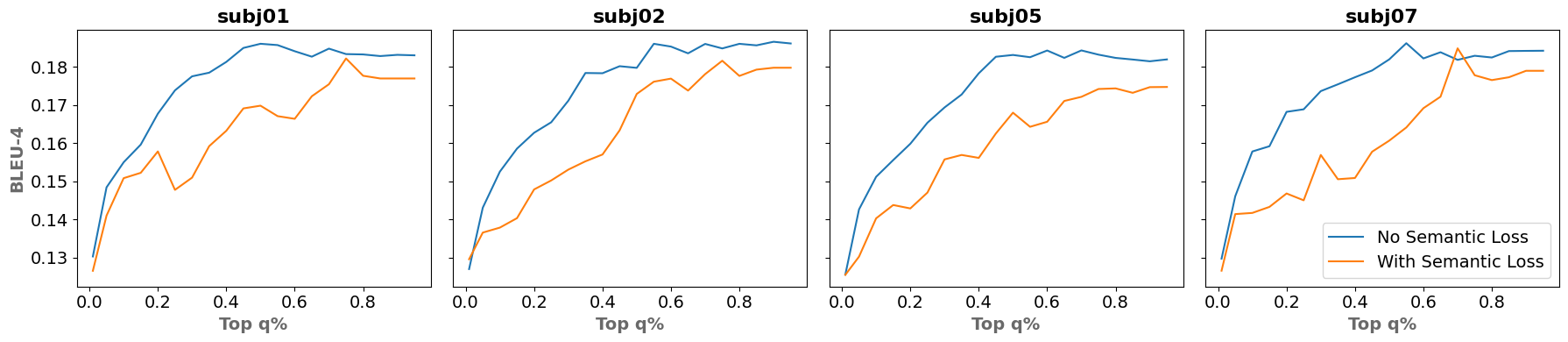}
  \vspace{-5mm}\caption{Performance comparison between test samples with and without semantic loss using n-gram metrics (BLEU-1, BLEU-2, BLEU-3, and BLEU-4).}\label{fig:loss_cap_ngram} \vspace{-5mm}
\end{figure*}

\begin{figure*}[t]
  \centering
  \includegraphics[width=\textwidth]{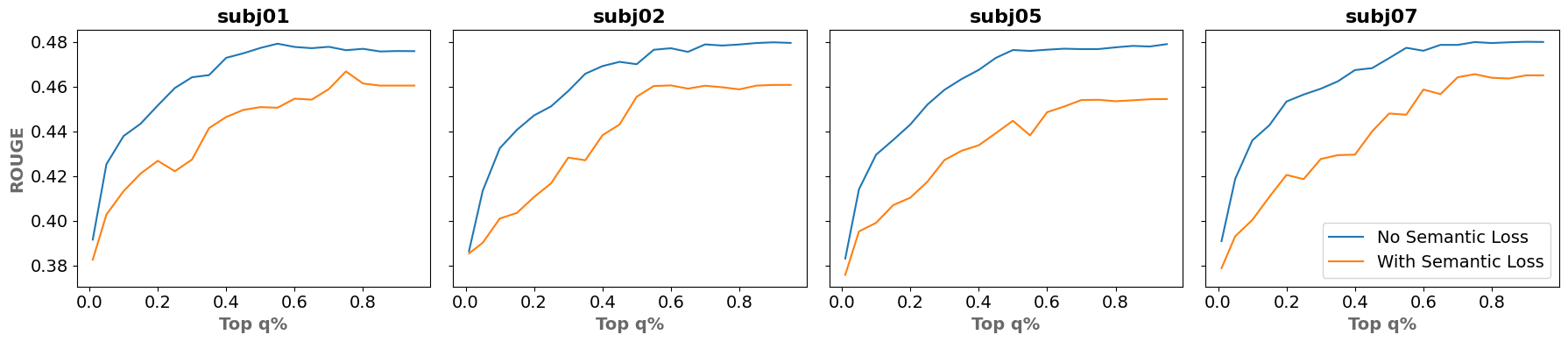}
  \includegraphics[width=\textwidth]{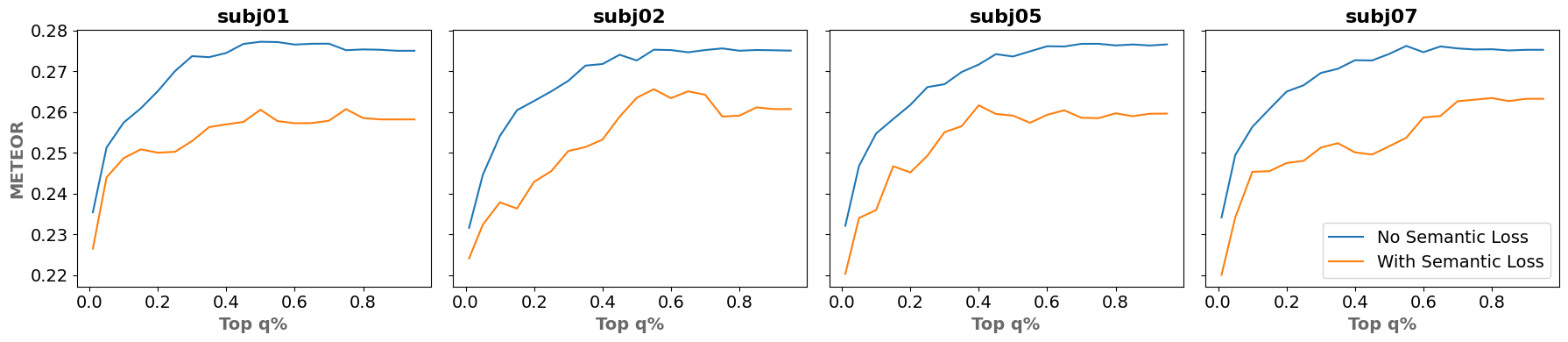}
  \includegraphics[width=\textwidth]{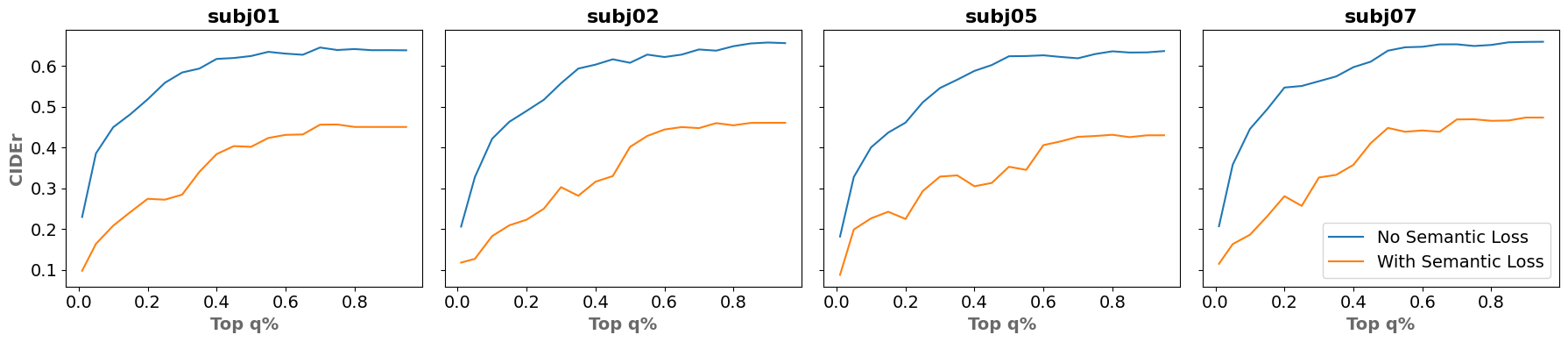}
  \includegraphics[width=\textwidth]{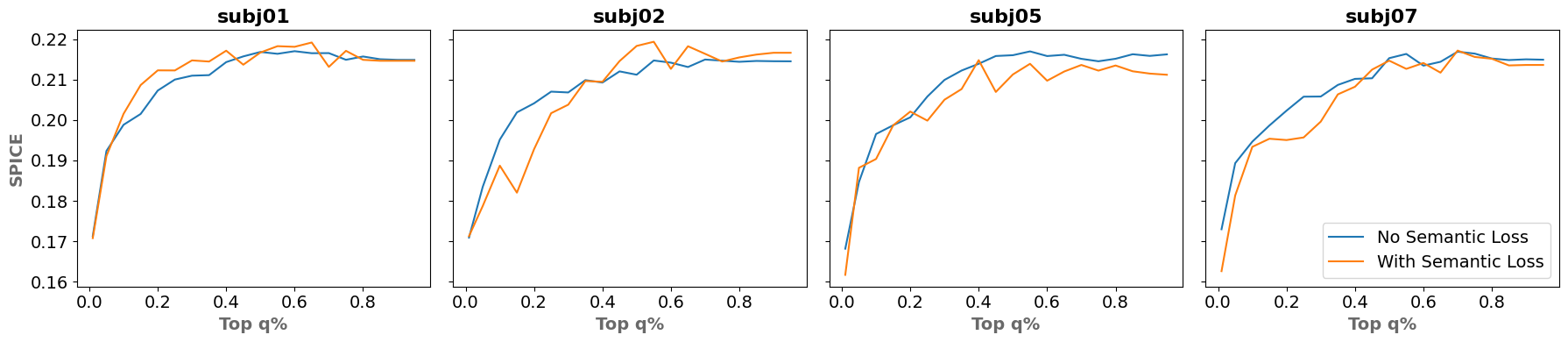}
  \vspace{-5mm}\caption{Performance comparison between test samples with and without semantic loss using semantic metrics (ROUGE, METEOR, CIDEr, and SPICE).}\label{fig:loss_cap_semantic} \vspace{-5mm}
\end{figure*}

\clearpage

\subsection{Captioning by subject-specific perception}\label{subsec:vis_more}
In Figure \ref{fig:vis_text} of the main paper, we show personalized caption generation results. Here, we provide more examples.

\begin{figure*}[h]
  \centering
  \includegraphics[width=\textwidth]{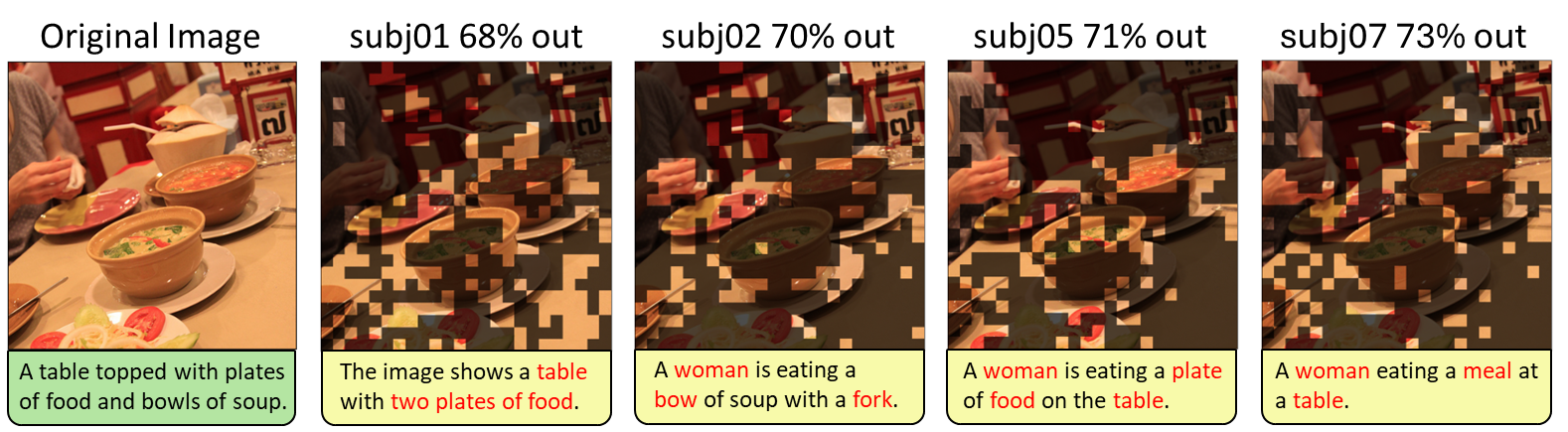}
  \includegraphics[width=\textwidth]{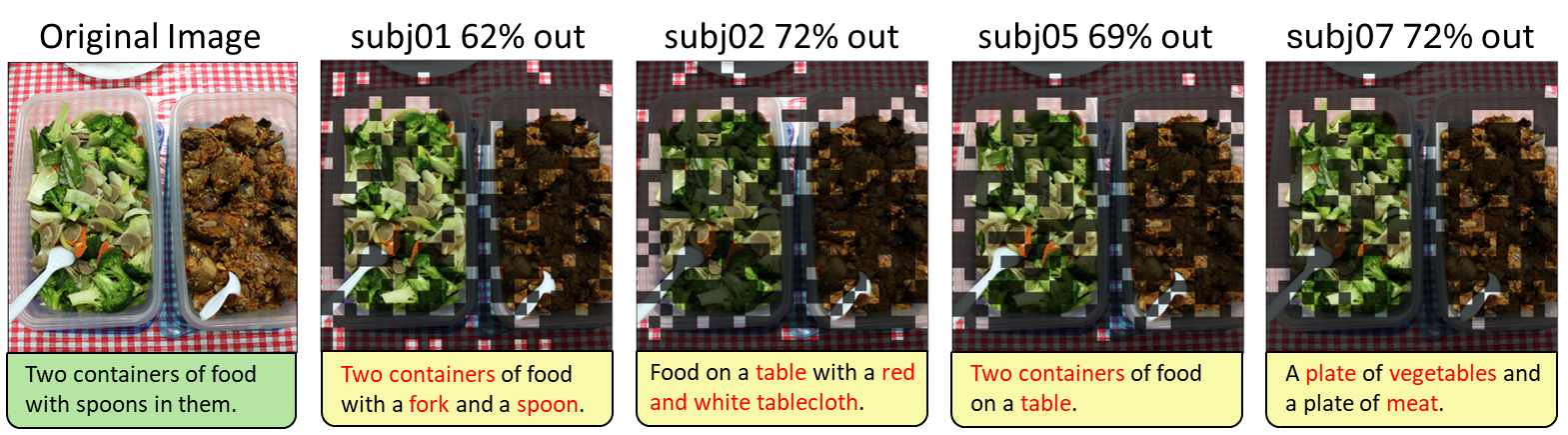}
  \includegraphics[width=\textwidth]{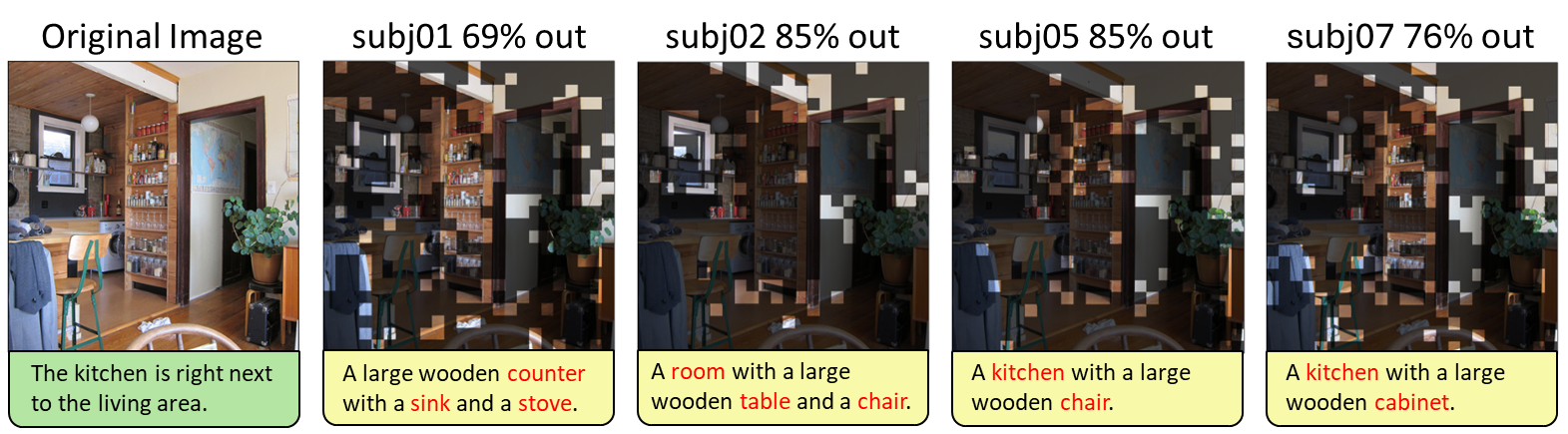}
  \includegraphics[width=\textwidth]{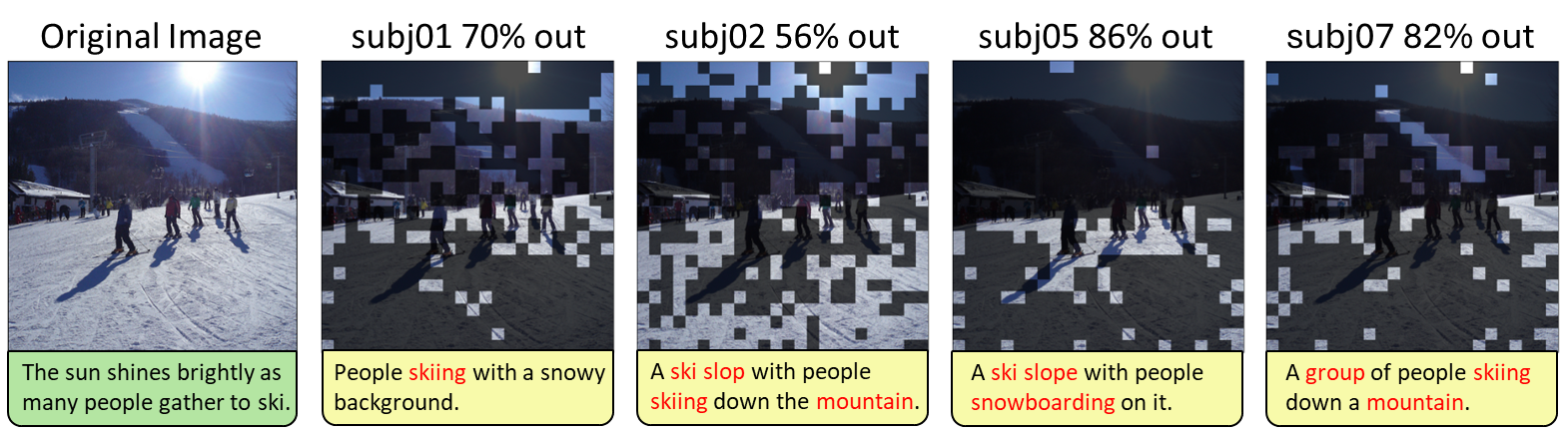}
  \vspace{-7mm}\caption{Various language generation and personalized visual perception, where we keep $5\%$ neural-visual activations for all subjects.}\label{fig:vis_more1} \vspace{-18mm}
\end{figure*}

\begin{figure*}[t]
  \centering
  \includegraphics[width=\textwidth]{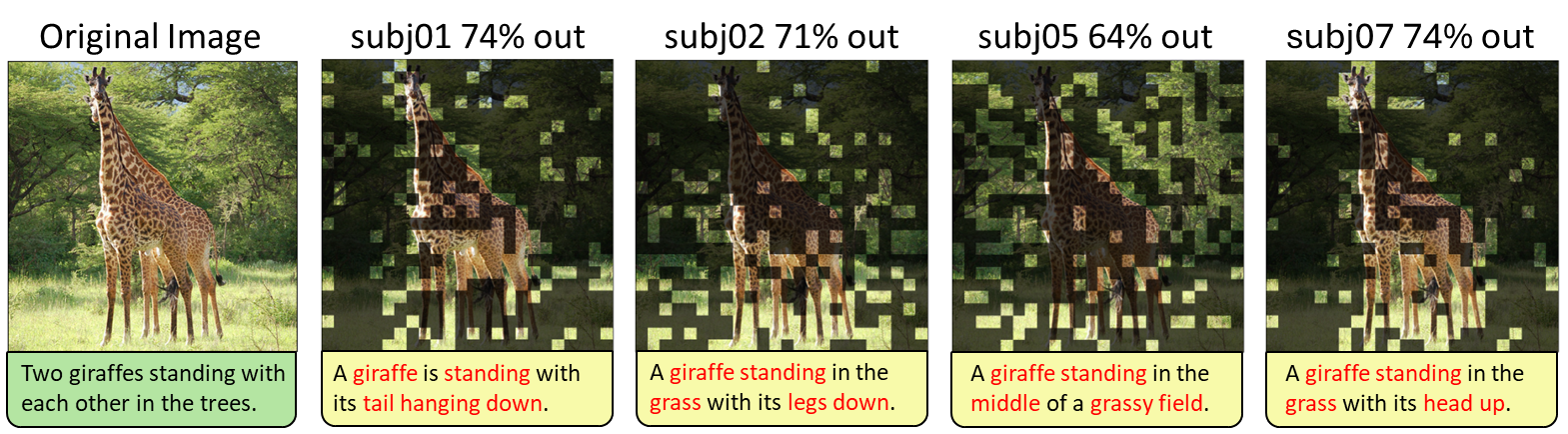}
  \includegraphics[width=\textwidth]{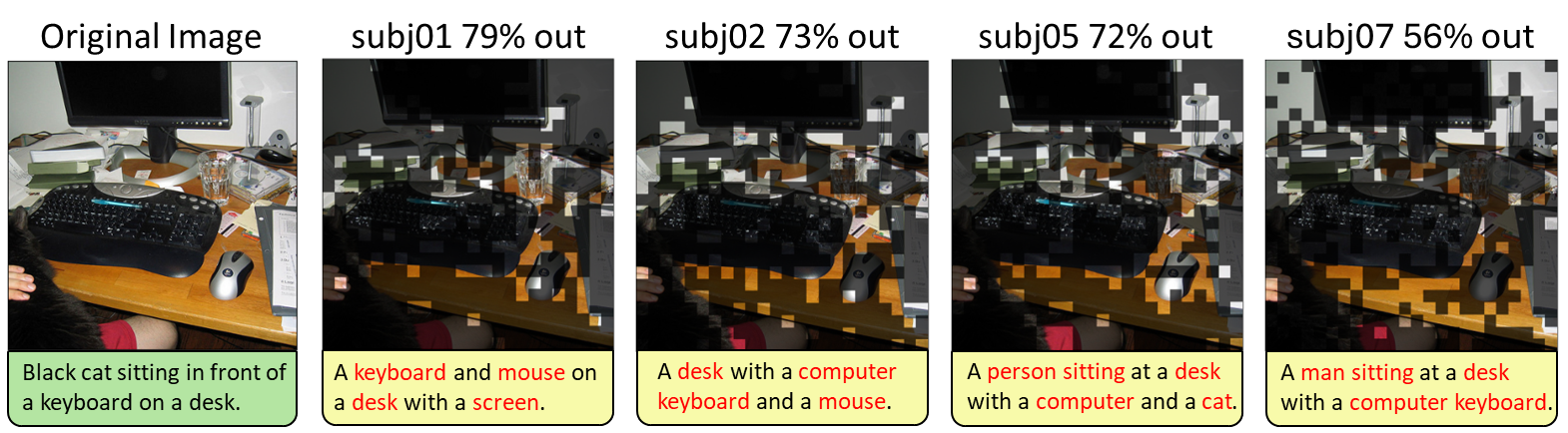}
  \includegraphics[width=\textwidth]{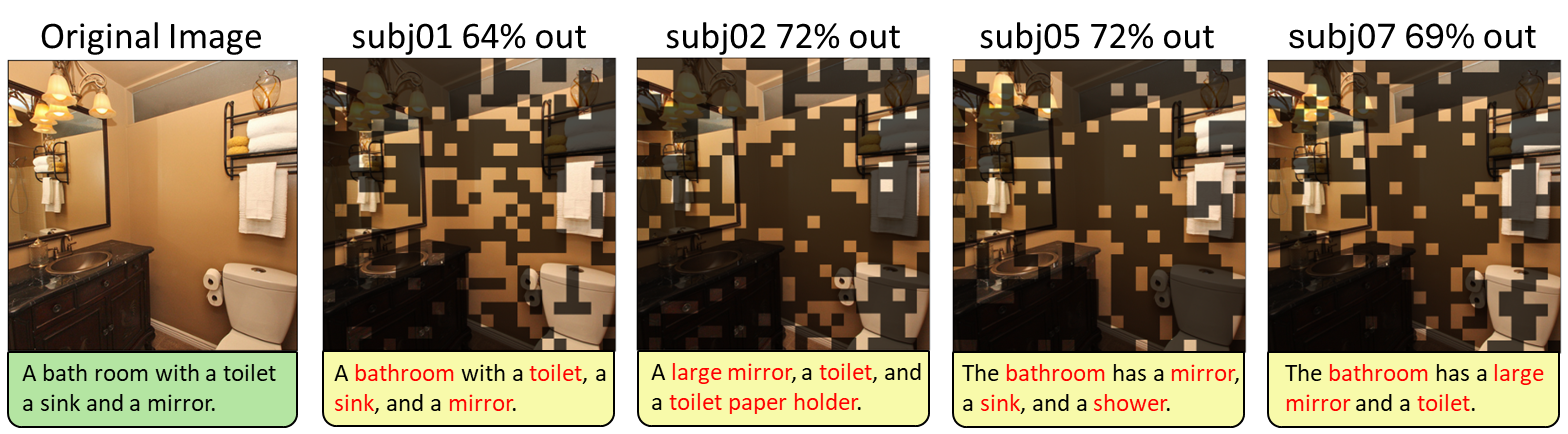}
  \includegraphics[width=\textwidth]{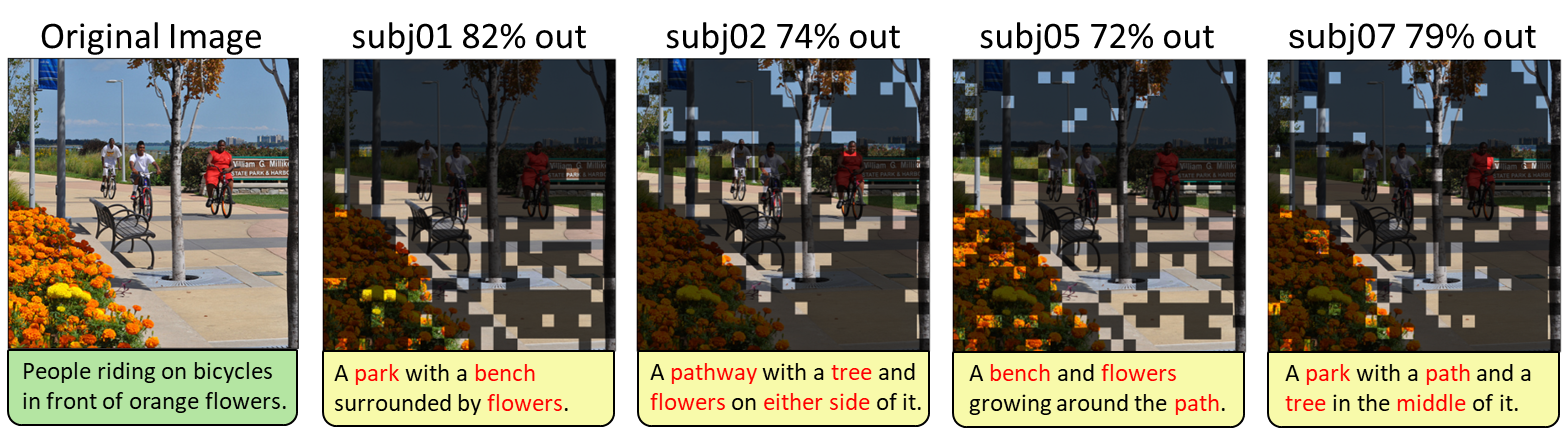}
  \vspace{-7mm}\caption{Various language generation and personalized visual perception, where we keep $5\%$ neural-visual activations for all subjects.}\label{fig:vis_more2} \vspace{-5mm}
\end{figure*}
\begin{figure*}[t]
  \centering
  \includegraphics[width=\textwidth]{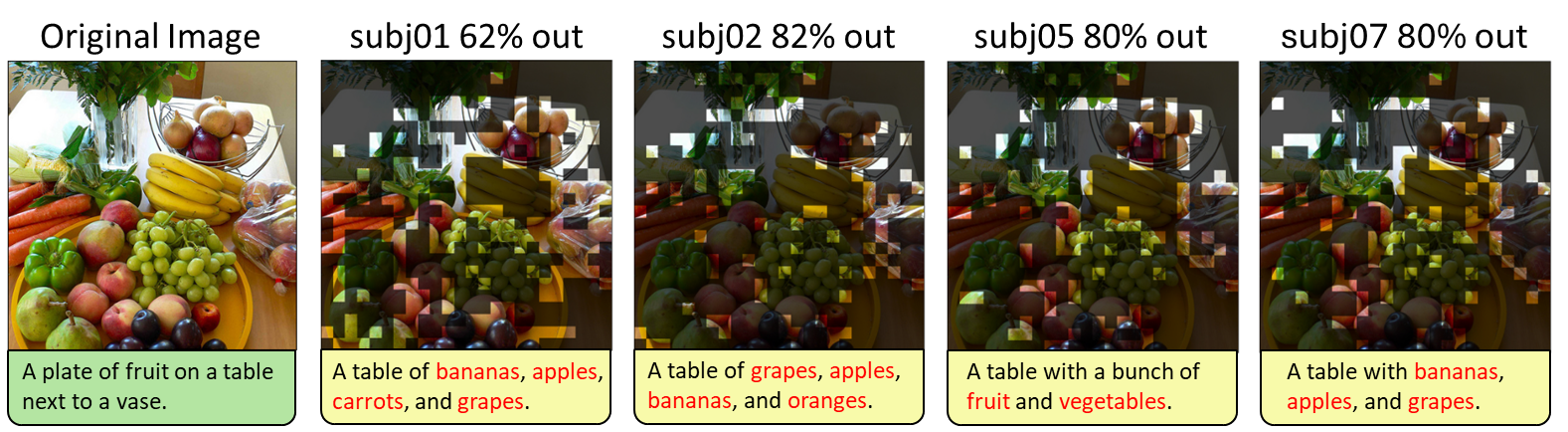}
  \includegraphics[width=\textwidth]{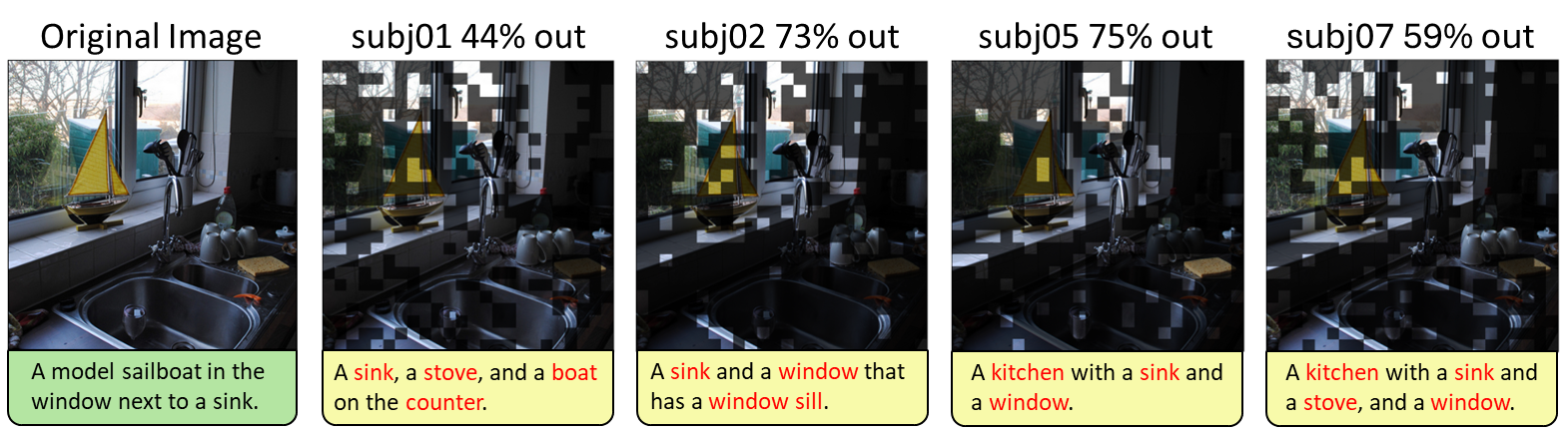}
  \includegraphics[width=\textwidth]{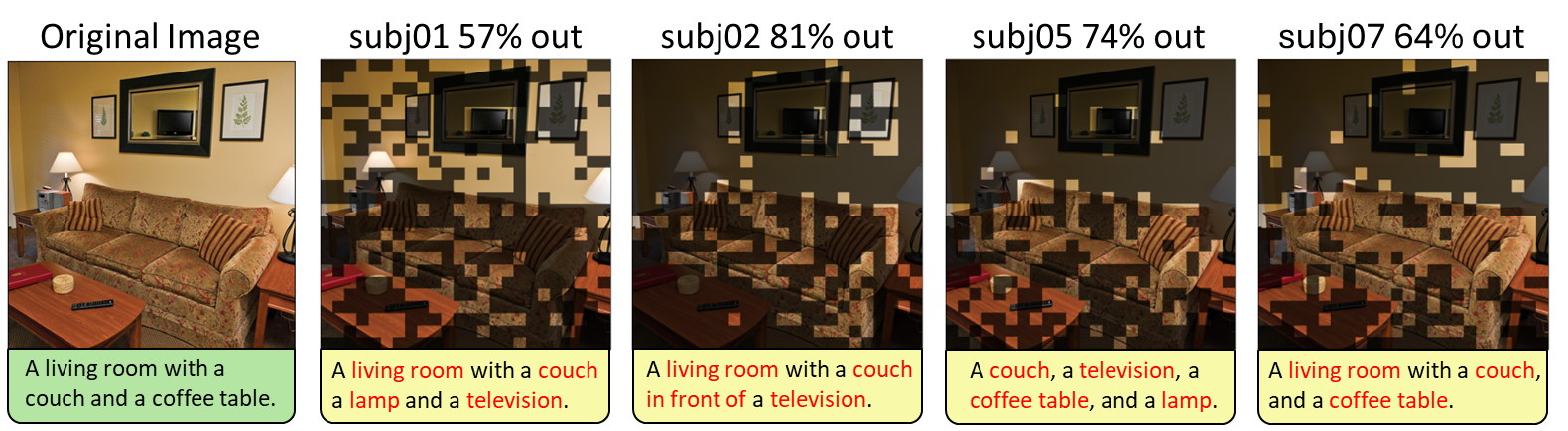}
  \includegraphics[width=\textwidth]{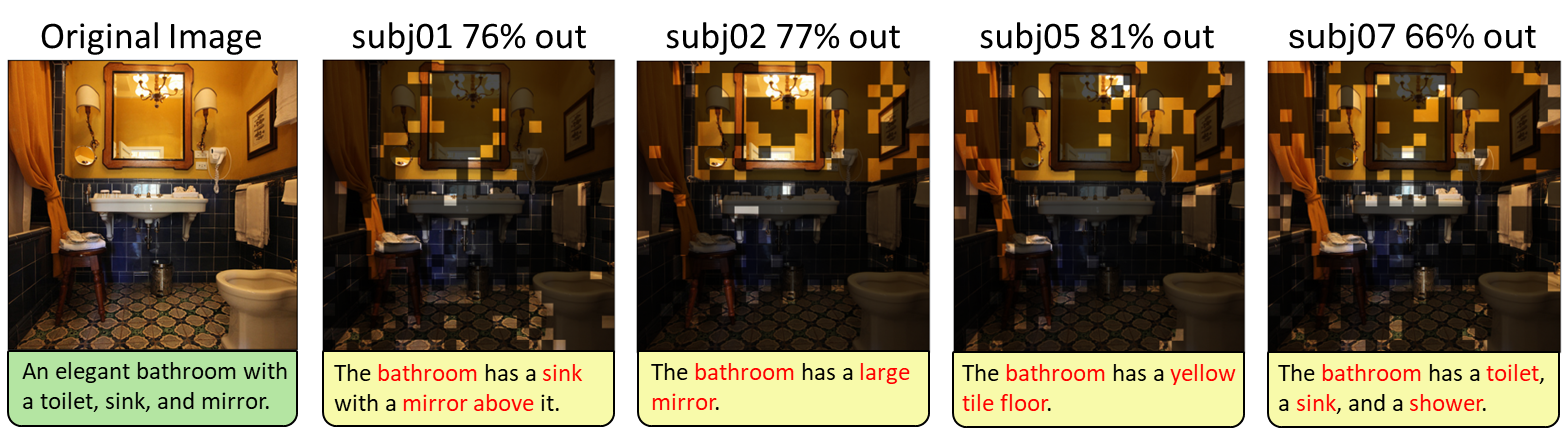}
  \vspace{-7mm}\caption{Various language generation and personalized visual perception, where we keep $5\%$ neural-visual activations for all subjects.}\label{fig:vis_more3} \vspace{-5mm}
\end{figure*}
\clearpage

\newpage
\section*{NeurIPS Paper Checklist}

\begin{enumerate}

\item {\bf Claims}
    \item[] Question: Do the main claims made in the abstract and introduction accurately reflect the paper's contributions and scope?
    \item[] Answer: \answerYes{} 
    \item[] Justification: we clearly state our contributions that match claims made in the abstract and introduction. We also conducted extensive experiments to validate our claims in the result section in the main paper and in the Appendix.

\item {\bf Limitations}
    \item[] Question: Does the paper discuss the limitations of the work performed by the authors?
    \item[] Answer: \answerYes{} 
    \item[] Justification: We discuss limitations in the Limitations paragraph of Section 5.

\item {\bf Theory assumptions and proofs}
    \item[] Question: For each theoretical result, does the paper provide the full set of assumptions and a complete (and correct) proof?
    \item[] Answer: \answerNA{} 
    \item[] Justification: we don't make any theoretical contributions in the paper.

    \item {\bf Experimental result reproducibility}
    \item[] Question: Does the paper fully disclose all the information needed to reproduce the main experimental results of the paper to the extent that it affects the main claims and/or conclusions of the paper (regardless of whether the code and data are provided or not)?
    \item[] Answer: \answerYes{} 
    \item[] Justification: All details necessary for reproducing the experimental results are fully described in either the main paper or the appendix. The code will be released upon acceptance.

\item {\bf Open access to data and code}
    \item[] Question: Does the paper provide open access to the data and code, with sufficient instructions to faithfully reproduce the main experimental results, as described in supplemental material?
    \item[] Answer: \answerYes{} 
    \item[] Justification: In the appendix, we describe the used dataset as well as the version of publicly available pretrained deep models we rely on in detail.

\item {\bf Experimental setting/details}
    \item[] Question: Does the paper specify all the training and test details (e.g., data splits, hyperparameters, how they were chosen, type of optimizer) necessary to understand the results?
    \item[] Answer: \answerYes{} 
    \item[] Justification: Please refer to the implementation details in the appendix for all the details regarding hyperparameter choice, preprocessing, etc.

\item {\bf Experiment statistical significance}
    \item[] Question: Does the paper report error bars suitably and correctly defined or other appropriate information about the statistical significance of the experiments?
    \item[] Answer: \answerNo{} 
    \item[] Justification: We don’t do a statistical significance check simply because our model is not directly comparable to conventional models. Also, the model is trained and tested on a large dataset. It would be time-consuming to do such tests.

\item {\bf Experiments compute resources}
    \item[] Question: For each experiment, does the paper provide sufficient information on the computer resources (type of compute workers, memory, time of execution) needed to reproduce the experiments?
    \item[] Answer: \answerYes{} 
    \item[] Justification: We list the computing resources needed for the proposed method, whose parameters are publicly available.
    
\item {\bf Code of ethics}
    \item[] Question: Does the research conducted in the paper conform, in every respect, with the NeurIPS Code of Ethics \url{https://neurips.cc/public/EthicsGuidelines}?
    \item[] Answer: \answerYes{} 
    \item[] Justification: We closely follow the code of ethics by NeurIPS.

\item {\bf Broader impacts}
    \item[] Question: Does the paper discuss both potential positive societal impacts and negative societal impacts of the work performed?
    \item[] Answer: \answerNA{} 
    \item[] Justification: the paper is an exploratory work and far from application. So, it may not
have any societal impacts for now.
    
\item {\bf Safeguards}
    \item[] Question: Does the paper describe safeguards that have been put in place for responsible release of data or models that have a high risk for misuse (e.g., pre-trained language models, image generators, or scraped datasets)?
    \item[] Answer: \answerNo{} 
    \item[] Justification: We don’t think the release of our data or models will have a high risk for misuse.

\item {\bf Licenses for existing assets}
    \item[] Question: Are the creators or original owners of assets (e.g., code, data, models), used in the paper, properly credited and are the license and terms of use explicitly mentioned and properly respected?
    \item[] Answer: \answerYes{} 
    \item[] Justification: the data and the pretrained deep models are fully described. We give credit to the original creator in the main paper.

\item {\bf New assets}
    \item[] Question: Are new assets introduced in the paper well documented and is the documentation provided alongside the assets?
    \item[] Answer: \answerYes{} 
    \item[] Justification: We will release our code upon acceptance, which includes new assets such as trained model weights

\item {\bf Crowdsourcing and research with human subjects}
    \item[] Question: For crowdsourcing experiments and research with human subjects, does the paper include the full text of instructions given to participants and screenshots, if applicable, as well as details about compensation (if any)? 
    \item[] Answer: \answerNo{} 
    \item[] Justification: Our research relies on human neural signal data, but it has no potential risks for participants.

\item {\bf Institutional review board (IRB) approvals or equivalent for research with human subjects}
    \item[] Question: Does the paper describe potential risks incurred by study participants, whether such risks were disclosed to the subjects, and whether Institutional Review Board (IRB) approvals (or an equivalent approval/review based on the requirements of your country or institution) were obtained?
    \item[] Answer: \answerNo{} 
    \item[] Justification: Our research relies on human neural signal data, but it has no potential risks for participants.

\item {\bf Declaration of LLM usage}
    \item[] Question: Does the paper describe the usage of LLMs if it is an important, original, or non-standard component of the core methods in this research? Note that if the LLM is used only for writing, editing, or formatting purposes and does \emph{not} impact the core methodology, scientific rigor, or originality of the research, declaration is not required.
    \item[] Answer: \answerNA{} 
    \item[] Justification: the core method development in this research does not involve LLMs as any
important, original, or non-standard components.

\end{enumerate}

\end{document}